\documentclass[10pt,letterpaper]{article}
\usepackage[top=0.85in,footskip=0.75in,marginparwidth=2in]{geometry}

\usepackage[utf8]{inputenc}

\usepackage{cite}

\usepackage{nameref,hyperref}

\usepackage[right]{lineno}

\usepackage{microtype}
\DisableLigatures[f]{encoding = *, family = * }

\raggedright
\usepackage{changepage}

\usepackage[aboveskip=1pt,labelfont=bf,labelsep=period,singlelinecheck=off]{caption}

\makeatletter
\renewcommand{\@biblabel}[1]{\quad#1.}
\makeatother

\usepackage{lastpage,fancyhdr,graphicx}
\usepackage{epstopdf}
\fancyheadoffset[L]{2.25in}
\fancyfootoffset[L]{2.25in}

\usepackage{color}

\definecolor{Gray}{gray}{.25}

\usepackage{graphicx}

\usepackage{sidecap}

\usepackage{wrapfig}
\usepackage[pscoord]{eso-pic}
\usepackage[fulladjust]{marginnote}
\reversemarginpar

\usepackage{amsmath}
\usepackage{subfig}
\usepackage{bm}
\newcommand{\etal}{{\em et al.}}
\newcommand{\incfig}{\centering\includegraphics}
\newcommand{\dd}[2]{\dfrac{d^2 #1}{d #2^2}} 

\let\baraccent=\= 
\renewcommand{\=}[1]{\stackrel{#1}{=}} 

\renewcommand{\dd}[2]{\ensuremath{\dfrac{\partial{#1}}{\partial{#2}}}}
\newcommand{\ddt}[2]{\ensuremath{\dfrac{\partial^2{#1}}{\partial{#2}^2}}}

\begin{document}
\vspace*{0.35in}

\begin{flushleft}
{\Large
\textbf\newline{Inverse Design of Single- and Multi-Rotor Horizontal Axis Wind Turbine
 Blades using Computational Fluid Dynamics}
}
\newline
\\
Behnam Moghadassian and
Anupam Sharma\textsuperscript{*}
\\
\bigskip
{\bf Department of Aerospace Engineering, Iowa State University, Ames, Iowa, 50011}
\\
\bigskip
* sharma@iastate.edu

\end{flushleft}

\section*{Abstract}
A method for inverse design of horizontal axis wind turbines (HAWTs) is
presented in this paper. The direct solver for aerodynamic analysis solves the
Reynolds Averaged Navier Stokes (RANS) equations, where the effect of the
turbine rotor is modeled as momentum sources using the actuator disk model
(ADM); this approach is referred to as RANS/ADM. The inverse problem is posed
as follows: for a given selection of airfoils, the objective is to find the
blade geometry (described as blade twist and chord distributions) which
realizes the desired turbine aerodynamic performance at the design point; the
desired performance is prescribed as angle of attack ($\alpha$) and axial
induction factor ($a$) distributions along the blade. An iterative approach is
used. An initial estimate of blade geometry is used with the direct solver
(RANS/ADM) to obtain $\alpha$ and $a$. The differences between the calculated
and desired values of $\alpha$ and $a$ are computed and a new estimate for the
blade geometry (chord and twist) is obtained via nonlinear least squares
regression using the Trust-Region-Reflective (TRF) method. This procedure is
continued until the difference between the calculated and the desired values is
within acceptable tolerance. The method is demonstrated for conventional,
single-rotor HAWTs and then extended to multi-rotor, specifically dual-rotor
wind turbines. The TRF method is also compared with the multi-dimensional
Newton iteration method and found to provide better convergence when
constraints are imposed in blade design, although faster convergence is
obtained with the Newton method for unconstrained optimization.


\section*{Introduction}
Rapid increase in utilization of turbines to harvest clean and renewable wind
energy resource has introduced new challenges for researchers. As power
production is directly dependent on the design of the wind turbine blades,
considerable research studies have focused on developing methods to improve
blade design in order to increase the output power. One approach to blade
design is to use direct analysis codes and perform parametric sweeps to
identify the highest-performing blade design. However, this approach is
computationally demanding and does not guarantee that the optimum design will
be reached~\cite{lee2015inverse}. More recently, researchers have started using
optimization algorithms in the blade design process. For example,
Chattot~\cite{chattot2003optimization} used the Lagrange multiplier
optimization method to maximize power while constraining thrust at a given tip
speed ratio. The inputs to the optimization algorithm were blade twist and
chord. A prescribed-wake vortex line method (VLM) was used that consisted of
the Goldstein model~\cite{goldstein1929vortex} to analyze the trailing wake
vortex structure behind the rotor blades and used the Biot-Savart formula to
calculate the induction.

Inverse algorithms were introduced in wind turbine blade design process by
Selig and Tangler~\cite{selig1995development}. They combined the
multi-dimensional Newton method with the Blade Element Momentum (BEM) theory to
perform inverse blade design.  This method is implemented in the software
PROPID which can perform inverse design in a variety of ways. One way is to
prescribe the desired distributions of axial induction factor ($a$) and lift
coefficient ($c_l$) along the blade, and the inverse design gives the blade
geometry (blade twist and chord distributions) that would yield the prescribed
$a$ and $c_l$ distributions. The desired radial distributions of $a$ and $c_l$
are carefully selected based on prior experience of the designer but with the
general aim of maximizing annual energy production (AEP).
Any combination of aerodynamic quantities, e.g., angle of attack
($\alpha$), lift coefficient ($c_l$), circulation ($\Gamma$), axial induction
($a$), etc. can be prescribed as desired distributions that the final blade
design is required to satisfy.

Lee~\cite{lee2015inverse} proposed the use of VLM over the BEM theory as the
direct solver in inverse design of HAWT blades to model the effects of
the three dimensional aerodynamic features of modern turbine blades, such as
sweep, pre-bend, non-planar wing tips, etc.  The VLM allows for partial
coupling of the different radial sections of the blade through induced velocity
computed via Biot Savart's law. Note that the coupling is partial because the
method still uses 2-D airfoil polars and cannot account for spanwise flow over
the blade. Radial distributions of $a$ and $c_l$ were used to prescribe the
desired blade aerodynamics. The multi-dimensional Newton method was used to
iterate on blade geometry (twist and chord) to obtain the prescribed $a$ and
$c_l$ distributions.

Selig and Coverstone-Carroll~\cite{selig1996application} and Giguere and
Selig~\cite{giguere1997desirable} merged the optimization techniques with
inverse design approaches to find the blade geometry that would maximize AEP.
They utilized the BEM theory along with a genetic algorithm to reach their goal
of maximizing AEP through blade design. If blade design is sought with the sole
objective of maximizing the AEP without any constraints, the final blade design
would have extremely high solidity inboard. The inboard $c_l$ and $a$
distributions are therefore tailored during inverse design procedure to yield a
practical blade design that would satisfy design-, manufacturing-, and
transportation constraints. Lee~\etal~\cite{lee2010two} used this idea to
determine the blade shape by considering a target function that includes the
AEP as well as the costs for blade masters, mold sets, tooling and blade
production. They used the strip theory~\cite{adkins1994design} to perform the
inverse design, while their target was to minimize energy loss at the Betz
optimum condition.  The inverse blade design was a part of a global
optimization algorithm which aimed to maximize power production and
simultaneously minimize blade cost.

Aerodynamics models that solve the Navier-Stokes equations offer higher
fidelity in analysis and design of wind turbine blades in comparison with
models based on solving the simplified potential flow equations (e.g., BEM and
VLM). The Navier-Stokes equations are usually solved with some simplifying
assumptions. For wind turbine aerodynamics application, the incompressible
turbulent flow (at least turbulence at high wavenumbers) is modeled (e.g.,
using eddy-viscosity based turbulence models), rather than directly solved.  If
the interest is only in mean quantities, time variation is ignored and Reynolds
Averaged Navier-Stokes (RANS) equations are used with appropriate turbulence
models that model the entire turbulence spectra. Many researchers
\cite{bachant2016modeling,lam2016study,alaimo20153d,marsh2015three,selvaraj2014numerical}
have utilized different forms of the RANS model to investigate wind turbines
aerodynamics. In one computationally efficient but simplified approach, the
blade geometry is not resolved in the simulation, instead, the effect of the
spinning blades on the air flow is introduced as source terms (typically as
body forces) in the Navier-Stokes equations.

The Actuator Line Method (ALM)~\cite{troldberg2008actuator} and the Actuator
Disk Method (ADM) \cite{sorensen1992unsteady} are the most commonly used
methods to simulate the effect of rotor blades using body forces without
resolving the blade geometry. Both methods use look-up tables for 2-D lift and
drag polars (obtained via prior experiments or simulations) with the local flow
velocity vector to calculate the net sectional force at each radially
discretized element of the blade. The net force is then applied as a spatially
distributed (typically a Gaussian distribution is used) source around the
radial element. Even though these methods use the strip theory approach with
2-D airfoil polars, they are still advantageous over potential flow methods in
that they can, to some degree, simulate turbulent wake mixing.  Inverse design
approaches that use such CFD techniques then have the potential to explore
designs that maximize wake mixing, and hence minimize wake losses in wind
farms. This paper presents a methodology to use RANS CFD to perform inverse
design of HAWT blades. The approach adopted here for inverse design is to
specify the desired $\alpha$ and $a$ distributions and let the inverse design
compute the blade twist and chord. The Trust-Region-Reflective method
\cite{byrd2000trust,coleman1994convergence} is used as the optimization
algorithm. The proposed inverse design procedure is verified for a number of
conventional, single-rotor horizontal axis wind turbines.

The inverse design method is then extended to multi-rotor wind turbines,
specifically the dual-rotor wind turbine (DRWT) proposed by Rosenberg
\etal~\cite{rosenberg2014novel}. Manufacturing and transportation constraints
on rotor blades of conventional, utility scale wind turbines result in
aerodynamically sub-optimal design in the blade root region (inner
25\%)~\cite{rosenberg2014novel}. This results in inefficient extraction of
energy by the rotor in this region. The
DRWT~\cite{rosenberg2014novel,mogh2016energies,rosenberg2016prescribed} uses a
secondary smaller rotor in front of the main rotor to efficiently harness the
energy from the airflow passing through the blade root region. Additionally,
the DRWT can enhance flow entrainment in the turbine wake, leading to faster
wake mixing and reduced wake
losses~\cite{moghadassian2015numerical,mogh2016energies,wang2016}.  It should
be emphasized that the inverse design methodology presented in this paper is
general enough to readily work for turbines with arbitrary number of rotor
disks and is not limited to DRWTs.

\section*{Computational Model}

\subsection*{Flow Solver}
\label{sec:adisk}
%
The incompressible, Reynolds Averaged Navier-Stokes (RANS) equations are solved
using the semi-implicit method for pressure linked equations (SIMPLE)
algorithm~\cite{patankar1972calculation}. The governing equations are
\begin{align}
 \dd{{\bar u_i}}{x_i} &= 0, \; {\rm and}, \nonumber \\
 {\bar u_j} \dd{ {\bar u_i}}{x_j} &=
-\dfrac{1}{\rho} \dd{\bar p}{x_i} - \nu \ddt{{\bar u_i}}{x_j} -  \dd{\overline{u'_i u'_j}}{x_j} + \dfrac{f_i}{\rho}.
 \label{eq:geq}
\end{align}
The two-equation $k-\epsilon$ turbulence model by Launder and
Spalding\cite{launder1974numerical}, with modifications suggested in Hargreaves
and Wright\cite{hargreaves2007use}, is used for turbulence closure. The source
term $f_i$ in the momentum equation, Eq.~(\ref{eq:geq}) is used to model the
force exerted by the turbine rotor blades on the fluid. This force is
determined using the blade element theory, which requires local flow velocity
at turbine location, geometric information about rotor blades such as chord and
twist, and blade aerodynamic characteristics. Blade aerodynamic
characteristics are specified as sectional lift and drag coefficients (airfoil
polars), and are provided by the user as look-up tables. The airfoil polars may
be corrected for rotational and dynamic stall effects. The body force, $f_i$ is
calculated using the Actuator Disk Method (ADM) by distributing the force over
a disk surrounding the turbine rotor using a Gaussian
distribution~\cite{mikkelsen2003actuator}. The numerical analysis procedure has
been implemented in the unstructured, finite volume solver simpleFOAM (part of
OpenFOAM) and has been previously validated against experimental
data~\cite{rosenberg2014novel,selvaraj2014numerical,rosenberg2016thesis}.

\subsection*{Simulation Set-up}
\label{sec:setup}

The objective of the paper is to demonstrate an inverse blade design procedure
that uses computational fluid dynamics (CFD) for aerodynamic analysis. To
minimize design time, a cost-effective computational setup is selected that is
consistent with traditional inverse design methods. Specifically, a uniform,
smooth inflow is considered, all rotor blades are assumed to be identical, and
only mean (time-steady) performance is investigated. With these approximations,
the problem becomes axisymmetric and makes the large number of calculations, as
required by inverse design procedures, conducive.

\begin{figure}[htb!]
  \incfig[width=0.75\columnwidth]{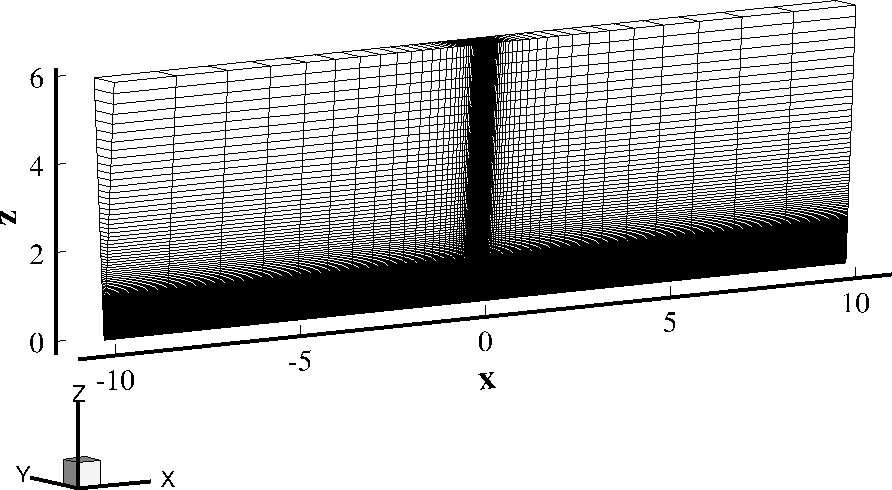}
  \caption{Isometric view of the computational mesh used for the proposed
  inverse design.}
  \label{fig:domain}
\end{figure}

Figure~\ref{fig:domain} shows an isometric view of an example axisymmetric mesh
used in the study with the $x$ axis as the symmetry axis. The radius of the
disk is along the $z$-direction. The lengths are nondimensionalized by the tip
radius of the turbine rotor, $R$. The domain is $20\,R$ in the $x$ direction
and $6\,R$ in the $z$ direction. It is one cell thick in the the $y$ direction
and the angle between the side planes is selected to be $1^{\circ}$. The
turbine rotor (or two rotors for a DRWT) is located around $x=0$ and the mesh
is refined in the region of the rotor disk to capture the large gradients
expected there; the mesh is also refined at the radial location corresponding
to the blade tip position ($z=1$).

The inflow is in the positive $x$ direction and the force exerted by the
turbine on the fluid is applied in the negative $x$ direction. For a rotor
spinning clockwise, as viewed from upstream, the torque force on the fluid is
applied in the positive $y$ direction. Axisymmetry boundary condition is
applied on the side planes. Zero gradient is imposed for pressure and velocity
at the outlet boundary. At the inlet, a zero-gradient pressure and a
fixed-value velocity boundary conditions are prescribed. The same setup was
used in previous
studies~\cite{rosenberg2014novel,selvaraj2014numerical,rosenberg2016thesis}
where the methodology was verified against experimental data for aerodynamic
performance prediction.

\subsubsection*{Mesh Sensitivity Study}
\label{sec:grid-independence}

Sensitivity of the RANS/ADM solver to mesh size is investigated with four
different meshes. Table~\ref{tab:grid} lists the mesh dimensions and the
corresponding computed turbine power coefficient, $C_P = 2\,P/ \rho u^3_\infty
\pi R^2$, where $P$ is the power extracted by the turbine, $\rho$ is air
density, $u_\infty$ is freestream flow speed, and $R$ is turbine rotor tip
radius.

\begin{figure}[htb!]
  \subfloat[angle of attack]        {\incfig[width=0.44\columnwidth]{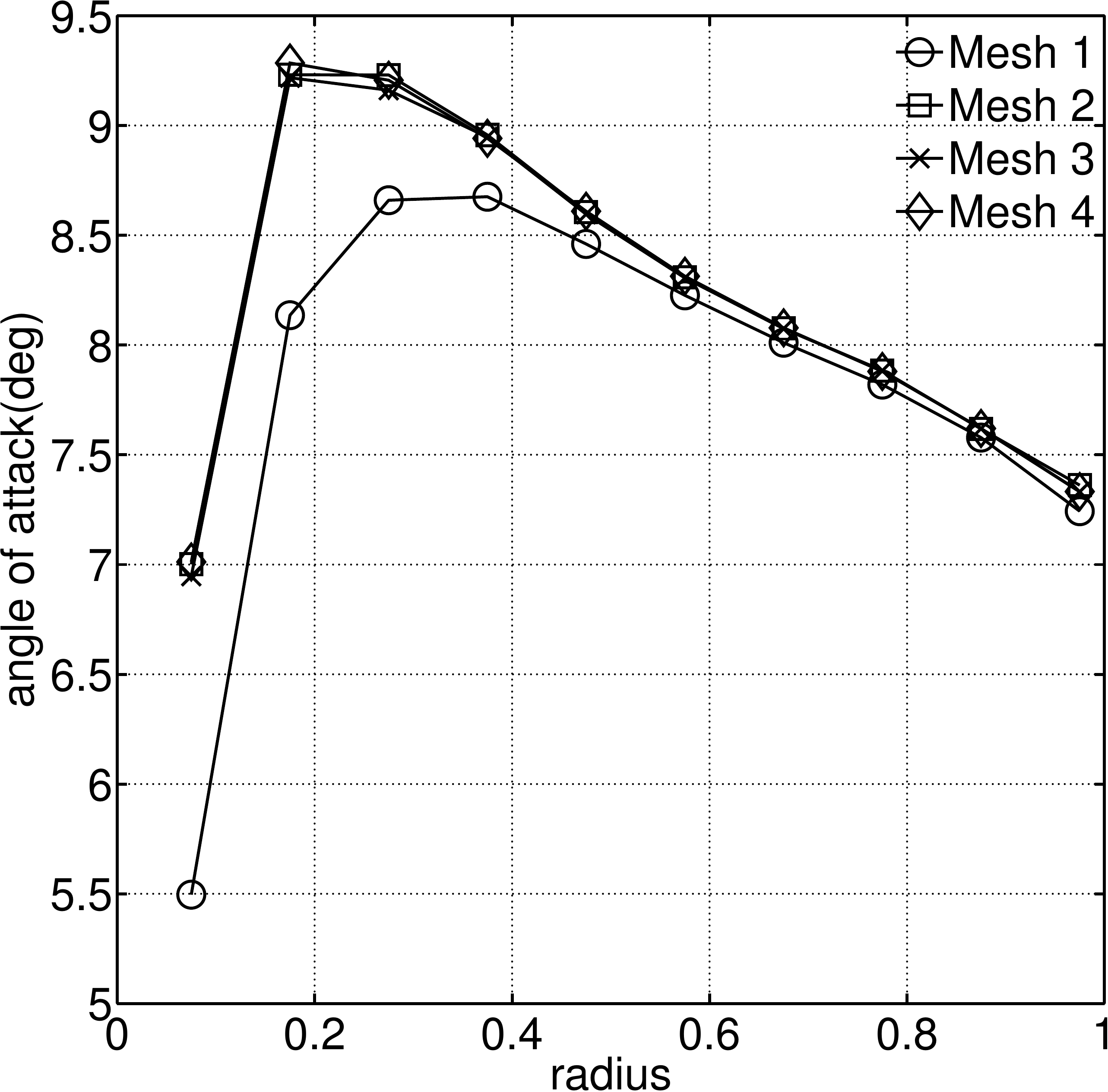}} \qquad
  \subfloat[axial induction factor] {\incfig[width=0.45\columnwidth]{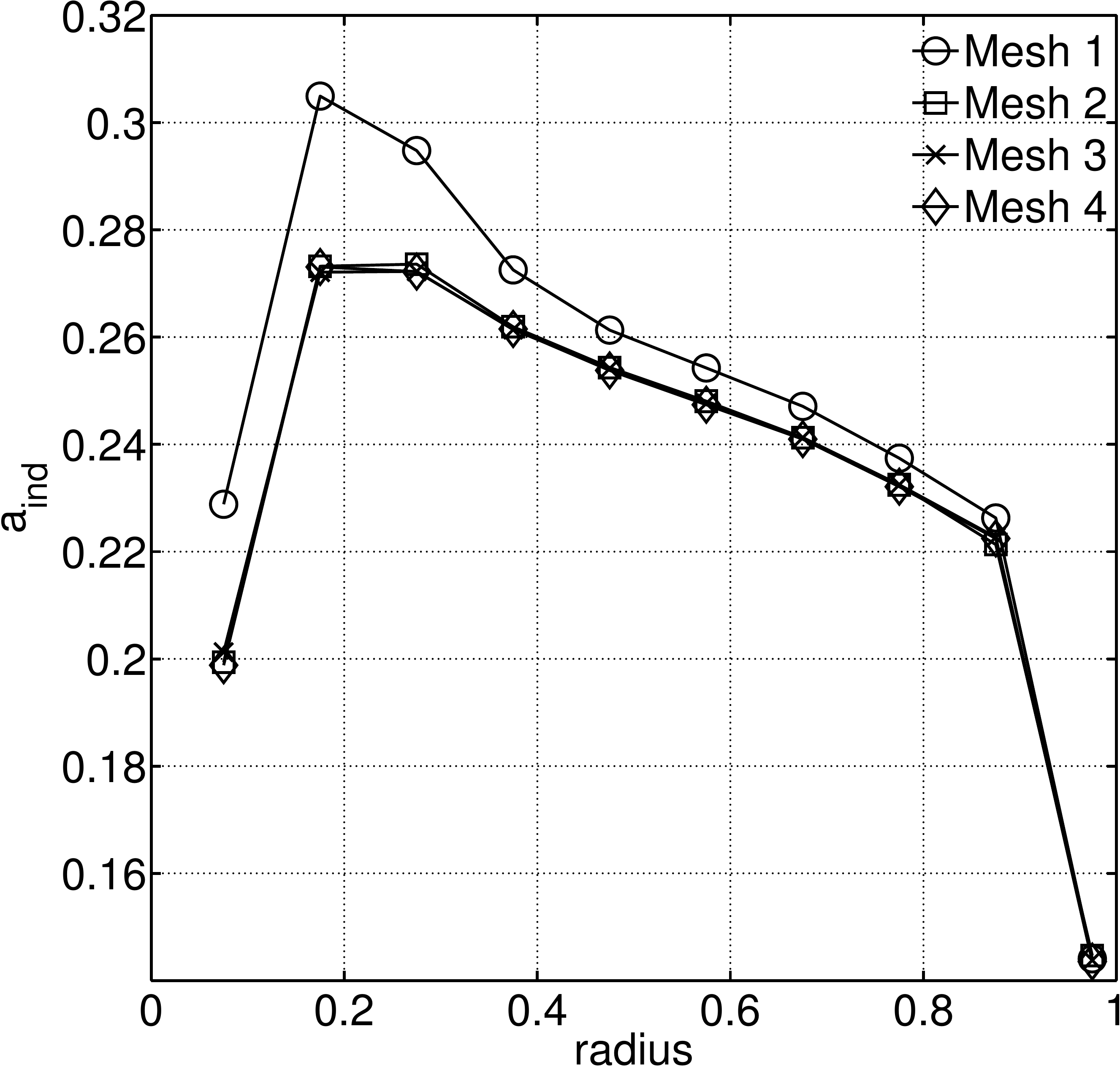}}
  \caption{Results of the mesh sensitivity study. RANS/ADM predicted
  distributions of angle of attack ($\alpha$) and axial induction factor ($a$)
  are compared for four different mesh sizes.}%
  \label{fig:meshSensitivity}%
\end{figure}

Variation with mesh size of radial distributions of angle of attack ($\alpha$)
and axial induction factor ($a$) are plotted in Fig.~\ref{fig:meshSensitivity}.
As seen in the figure, refining the grid beyond $N_x \times N_z=101 \times 229$
(Mesh 2) does not adversely impact the results; the $C_P$ becomes constant and
the variations in radial distributions are minimal. Hence, the grid resolution
of Mesh 2 is selected to obtain the results presented in this paper. This study
is conducted for a conventional single-rotor wind turbine and grids for
multi-rotor wind turbines are deduced from this assessment. A detailed mesh
sensitivity study for this solver has been reported previously in
Ref.~\cite{thelen2016direct}.

\begin{table}[ht]
 \caption{Mesh sensitivity study: grid dimensions and aerodynamic power
 coefficient.}
 \label{tab:grid}
 \begin{center}
   \begin{tabular}{c|c|c}
     \hline \hline
     Mesh \# & $N_x \times N_z$ & $C_P$ \\
     \hline \hline
     1 & $79  \times 141$ & $0.483$ \\
     2 & $101 \times 229$ & $0.485$ \\
     3 & $117 \times 279$ & $0.485$ \\
     4 & $136 \times 329$ & $0.485$ \\
     \hline \hline
   \end{tabular}
 \end{center}
\end{table}

\subsection*{Solution Algorithm}
\label{sec:algorithm}

The purpose of using an inverse scheme in a wind turbine blade design process
is to directly compute blade geometries which give the desired aerodynamic
performance. The designer can choose the target aerodynamic performance
parameters and prescribe their desired values (distributions). If the
prescribed values of the aerodynamic performance parameters are physically
consistent and achievable, the inverse design process should give the turbine
blade geometry that meets the desired performance.

While there are multiple ways in which the inverse design problem can be posed,
we choose to prescribe distributions of axial induction factor ($a$), and angle
of attack ($\alpha$) as desired outcomes. As an example, in an ideal
single-rotor turbine, the desired $a$ could be $1/3$ based on the 1-D momentum
theory, and the target $\alpha$ values could be selected to give the highest
lift-to-drag ratio for the airfoil at that radial location. In other studies,
$c_l$ distributions, as opposed to $\alpha$ distributions have been
prescribed~\cite{lee2015inverse}. There is one potential problem with that
choice -- the $\alpha-c_l$ curve is multi-valued, i.e., a given value of $c_l$
can occur for multiple $\alpha$ values (pre- and post-stall). While at a
typical design point, all radial locations are in the pre-stall region, an
iterative design procedure can have transients in the post-stall region and
hence the duality of $\alpha-c_l$ curve can pose problems.  Prescription of
$\alpha$ is still based on the desired $c_l/c_d$ (presumably where it is
maximum), but it ensures that the turbine is not designed to operate in the
post-stall regime of that airfoil. Therefore, in this study the desired
parameters are set to be $\alpha$ and $a$. The inverse design algorithm then
proceeds as follows:
\begin{enumerate}
  \item The blade is discretized along the span into $m$ segments and the
    desired values of $\alpha$ and $a$ are prescribed for each segment $j$ as a
    1-D vector ${\bf b}=\{\alpha_j,a_j\}^T$.

  \item For the first iteration ($n=0$), an initial guess of the blade geometry
    is provided as a 1-D vector, ${\bf x}_0=\{c_j,\theta_j\}^T$ of length
    $2\,m$, where $c_j$ and $\theta_j$ are the blade chord and twist angle at
    segment $j$.

  \item The inverse subroutine is invoked (details provided in
    Sec.~\ref{sec:inverse}):

  \begin{enumerate}
    \item With ${\bf x}_n$ as the baseline design, the direct solver is called
      $2\,m+1$ times, to measure the effect of perturbing (by an infinitesimal
      amount) each element of ${\bf x}_n$ on ${\bf b}_n$ (distributions of
      $\alpha$ and $a$ at iteration $n$). The Jacobian (sensitivity) matrix
      ($=\partial\{\alpha,a\}/\partial\{\theta,c\}$, see
      Fig.~\ref{fig:jacobian}) is calculated using first order, forward finite
      differences.

    \item A new estimate of the blade geometry ${\bf x}_{n+1}$ is computed
      using the Jacobian matrix and the Trust-Region-Reflective method
      (described in Sec.~\ref{sec:inverse}) with the aim of minimizing the
      difference between the desired and the computed distributions of $\alpha$
      and $a$.

    \item The direct solver (RANS/ADM) is used to evaluate the aerodynamic
      performance of the new turbine geometry (${\bf x}_{n+1}$) and obtain
      ${\bf b}_{n+1}=\{\alpha_j,a_j\}^T$.

    \item If the ${\bm l}^2$ norm of the difference between the desired and
      computed aerodynamic performance ${\bf b}$ $(=\{\alpha_j,a_j\}^T$) is
      within the desired tolerance (this is just one of many stopping
      criteria), the algorithm is terminated and the current blade geometry
      ${\bf x}_{n+1}$ is output as the final blade design. Otherwise, the
      algorithm returns to step 3(a) and the iterations continue with the
      iteration counter $n$ incremented to $n+1$.
  \end{enumerate}
\end{enumerate}

\begin{figure}[htb!]
  \incfig[width=0.65\columnwidth]{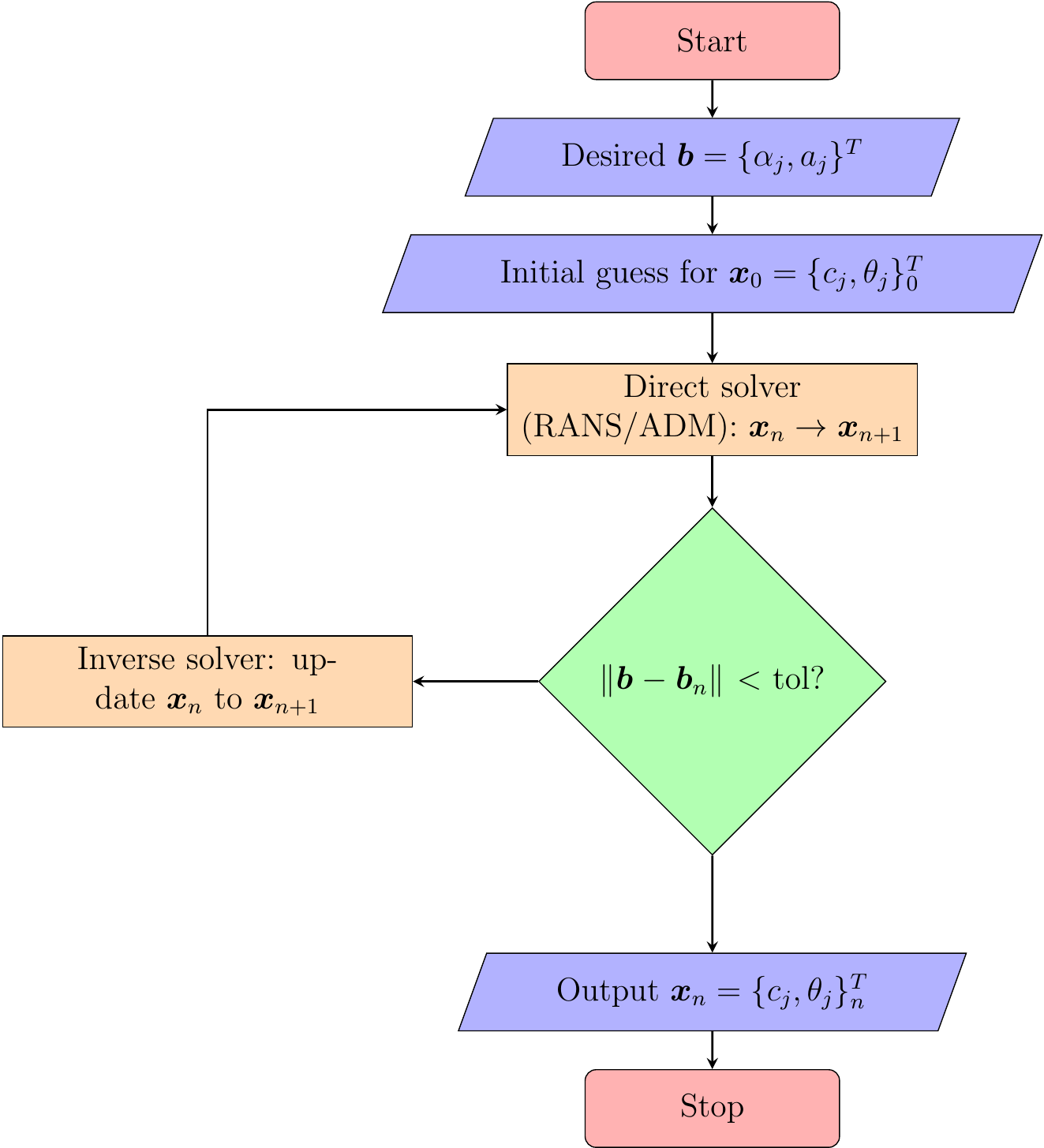}
  \caption{Flowchart of the inverse design algorithm.}%
  \label{fig:flowchart}%
\end{figure}

The algorithm is presented as a flowchart in Fig.~\ref{fig:flowchart}.

\subsection*{Inverse Solver}
\label{sec:inverse}
%
This section provides a brief introduction to the selected inverse solver, the
Trust-Region-Reflective (TRF) method.  Suppose the minimum of a function $f$ is
sought in a bounded or unbounded domain. As a first guess, a point $x_0$ in the
domain is randomly picked and the function $f$ is evaluated at $x_0$. The
essential idea behind this method is to approximate $f$ with a simpler function
$q$ (see Eq.~\ref{eq:min}) that behaves similarly to $f$ in the vicinity (trust
region, $N$) of $x_0$ and find a new point $x_1$ in $N$ where $q$ is minimum.
Now if the reduction in the approximated function ($q$) is also inherited by
the original function $f$, $x_1$ is accepted as an updated estimate of the
location of minimum value of $f$, and the procedure is repeated. Otherwise the
trust region ($N$) is reduced in size and the search for minimum $q$ is
repeated. In order to reduce the computational time, the trust-region
sub-problems are confined to two dimensional sub-spaces $S$. In fact, the
sub-problem is defined as
{\small{
\begin{equation}
 \arg\min \{q(s)\}, \text{ subject to } ||Ds||\le\Delta \text{ and }
s \in \text{span}[s^U_k,s^{FS}_k],
 \label{eq:min}
\end{equation}
}}
where $q(s) = f(x_k)+s^Tg+\frac{1}{2}s^THs$ is the approximation for $f$ around
$x_k$, $g$ is the gradient of $f$ at $x_k$, and $H$ is the Hessian matrix
(symmetric matrix of second derivatives of $f$), $D$ is the diagonal scaling matrix,
$\Delta$ is the radius of the trust region, $s^U_k$ is the steepest descent
direction given by $s^U_k = -g \, ({g^T g})/({g^T H g})$, and $s^{FS}_k$ is
either an approximate Newton direction, $H\cdot s^{FS}_k = -g$, or a direction
of negative curvature, $(s^{FS})^T_k\cdot H\cdot s^{FS}_k <0$.  This
formulation results in global convergence through the steepest descent or
negative curvature direction while achieving a fast local convergence, when it
exists, using the Newton step\cite{byrd1987trust,byrd2000trust} in each trust
region.

In this study, the multi-dimensional function $\bf F$ is defined as
\begin{equation}
 {\bf F}({\bf x}) = {\bf b}^d - {\bf b}_n^{c} = \left[\begin{matrix}
  F_1 ({\bf x})\\
  F_2 ({\bf x})\\
  \vdots \\
  F_{m} ({\bf x})\\
  F_{m+1} ({\bf x})\\
  \vdots \\
  F_{2\,m} ({\bf x}) \end{matrix}\right]
\;
  =
\;
  \left[\begin{matrix}
 \alpha_1^d - \alpha_1^c\\
  \alpha_2^d - \alpha_2^c\\
  \vdots \\
  \alpha_m^d - \alpha_m^c\\
  a_1^d - a_1^c\\
  \vdots \\
  a_m^d - a_m^c \end{matrix}\right],
  \label{e:function}
\end{equation}
where superscripts $d$ and $c$ stand for the desired and calculated values
respectively, and $\bf b$ is the vector of design parameters.  There are
multiple stopping criteria for this algorithm: maximum number of inverse
iterations, minimum values of the target function, magnitude of change in the
independent variables, and norm of gradient of the target function.  The
algorithm stops when any of these criteria is met.  Based on the preliminary
tests, the maximum number of inverse iterations is set to 30 and the minimum
values of $\lVert \mathbf{F} \lVert$, $\lVert \Delta\mathbf{x} \lVert$ and
$\lVert \Delta\mathbf{g} \lVert$ are set to $10^{-8}$.  The discussions on the
merits of the TRF method over other optimization schemes, details of how the
algorithm proceeds under different circumstances, and determination of the size
of domain $N$ for the TRF algorithm can be found in
Refs.~\cite{byrd1987trust,byrd2000trust}. The interested reader is referred to
Refs.~\cite{coleman1994convergence,byrd2000trust,byrd1987trust} for a detailed
description of the method.

\section*{Verification of Inverse Design Methodology}
\label{sec:verification}
%
The proposed inverse design procedure is tested for three conventional, single
rotor wind turbines (SRWTs) A description of these test cases and the
performance of the inverse design procedure are presented in this section. The
purpose of these tests is to ensure that the algorithm is capable of obtaining
blade designs which satisfy different design target values for blades made of
different airfoils. It should be noted that the direct solver (RANS/ADM model)
utilized in this study has been validated
previously~\cite{rosenberg2014novel,selvaraj2014numerical} for wind turbine
aerodynamic performance prediction. For optimization, the nonlinear
least-square solver package, {\it{scipy.optimize}}, available with the {\em
Python} scripting language is employed.

\subsection{Test Case $1$: Single-Rotor Betz Optimum Turbine}
\label{sec:test1}

As a first test case, we attempt to design the Betz optimum rotor. Per the 1-D
momentum theory, the axial induction factor should be $1/3$ over the entire
rotor disk to achieve maximum $C_P$. The turbine blade is desired to be
designed using one airfoil (the 18\% thick DU-96-W180 airfoil) for the entire
blade span. This airfoil has a high lift-to-drag ratio and typically is used
for tip sections of utility-scale turbine rotor
blades~\cite{rosenberg2014novel}. To achieve the best performance, the desired
$\alpha$ is selected to be 10 degrees, which is where $c_l/c_d$ is maximum for
the DU-96-W180 airfoil. Therefore, $\alpha_j=10^{\circ},\; a_j=1/3$ is
specified for all radial segments $\forall j \in [1,m]$. The design tip speed
ratio, $\lambda = \Omega \, R/u_\infty$ is set to be $7.0$.

The design algorithm needs to compute the Jacobian matrix
$\left(=\partial\{\alpha_j,a_j\}/\partial\{\theta_k,c_k\}\right)$ in order to
find the new minimum point at each iteration~\cite{byrd1987trust}. Information
about the Jacobian matrix is also useful to understand how output variables,
$\{\alpha_j,\, a_j\}$ at the $j^{th}$ radial segment of the blade, are
influenced by change in input variables $\{c_k,\,\theta_k\}$ at the $k^{th}$
radial location on the blade; $\forall j,k \in [1,m]$.

\begin{figure}[htb!]
  \centering
  \subfloat[Schematic]                        {\incfig[width=0.25\columnwidth]{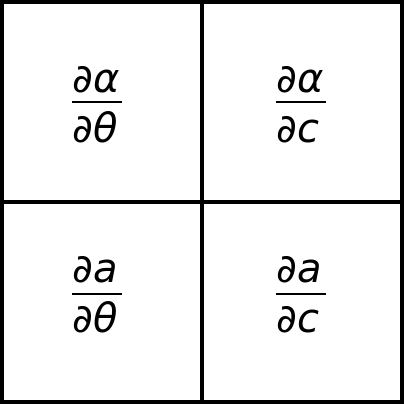}} \\
  \subfloat[$\partial \alpha/\partial \theta$]{\incfig[width=0.45\columnwidth]{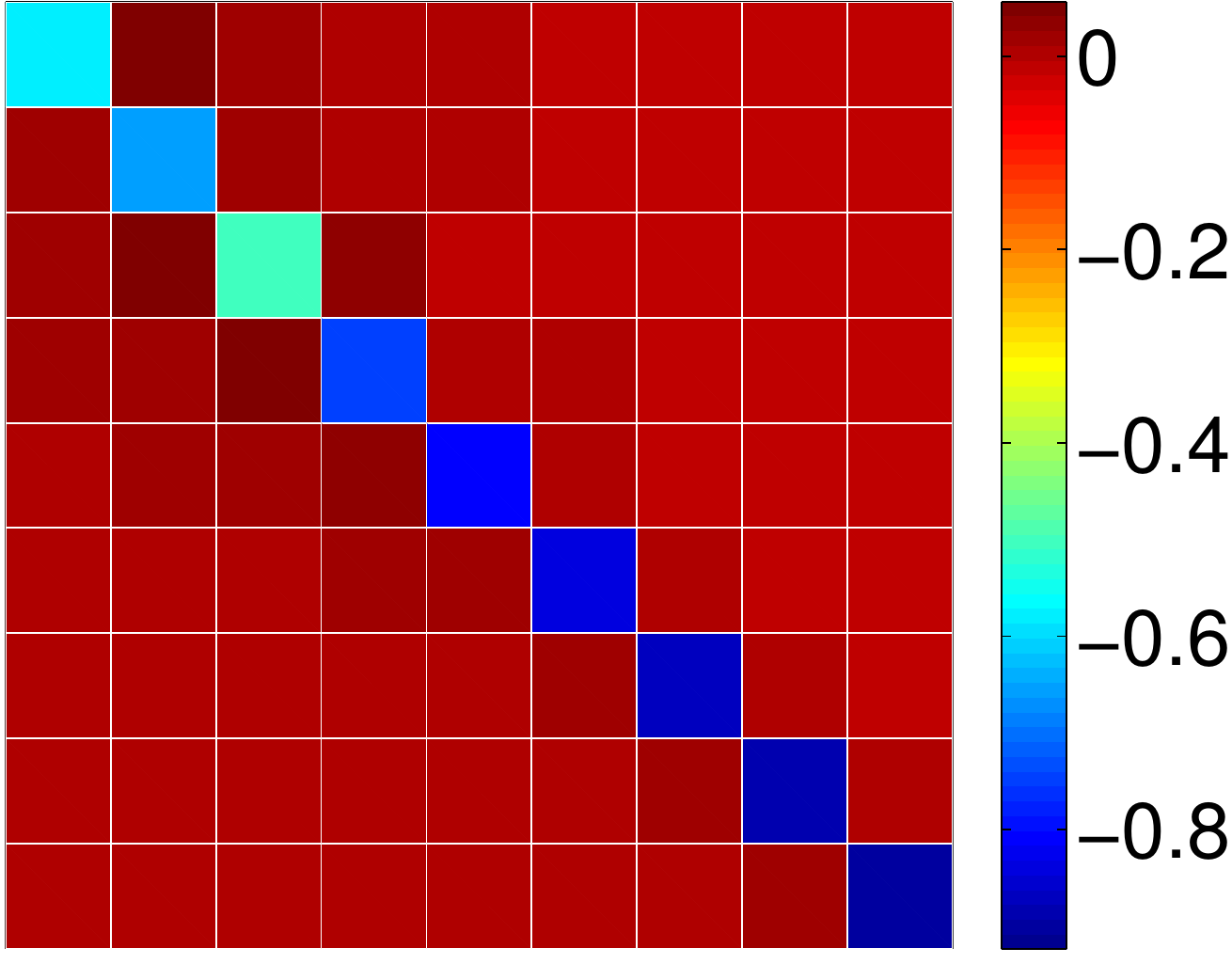}} \qquad
  \subfloat[$\partial \alpha/\partial c$     ]{\incfig[width=0.45\columnwidth]{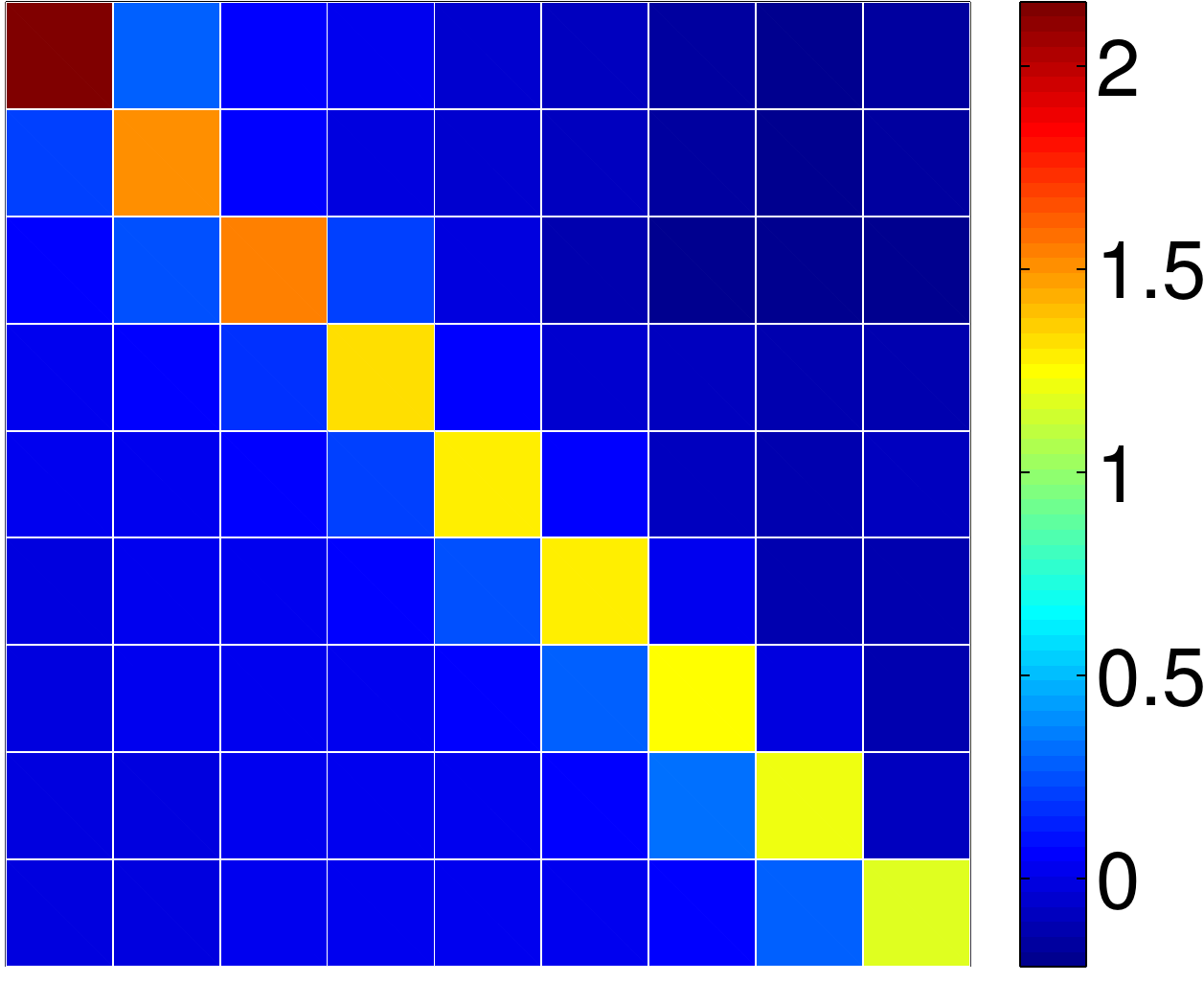}} \\
  \subfloat[$\partial  a    /\partial \theta$]{\incfig[width=0.45\columnwidth]{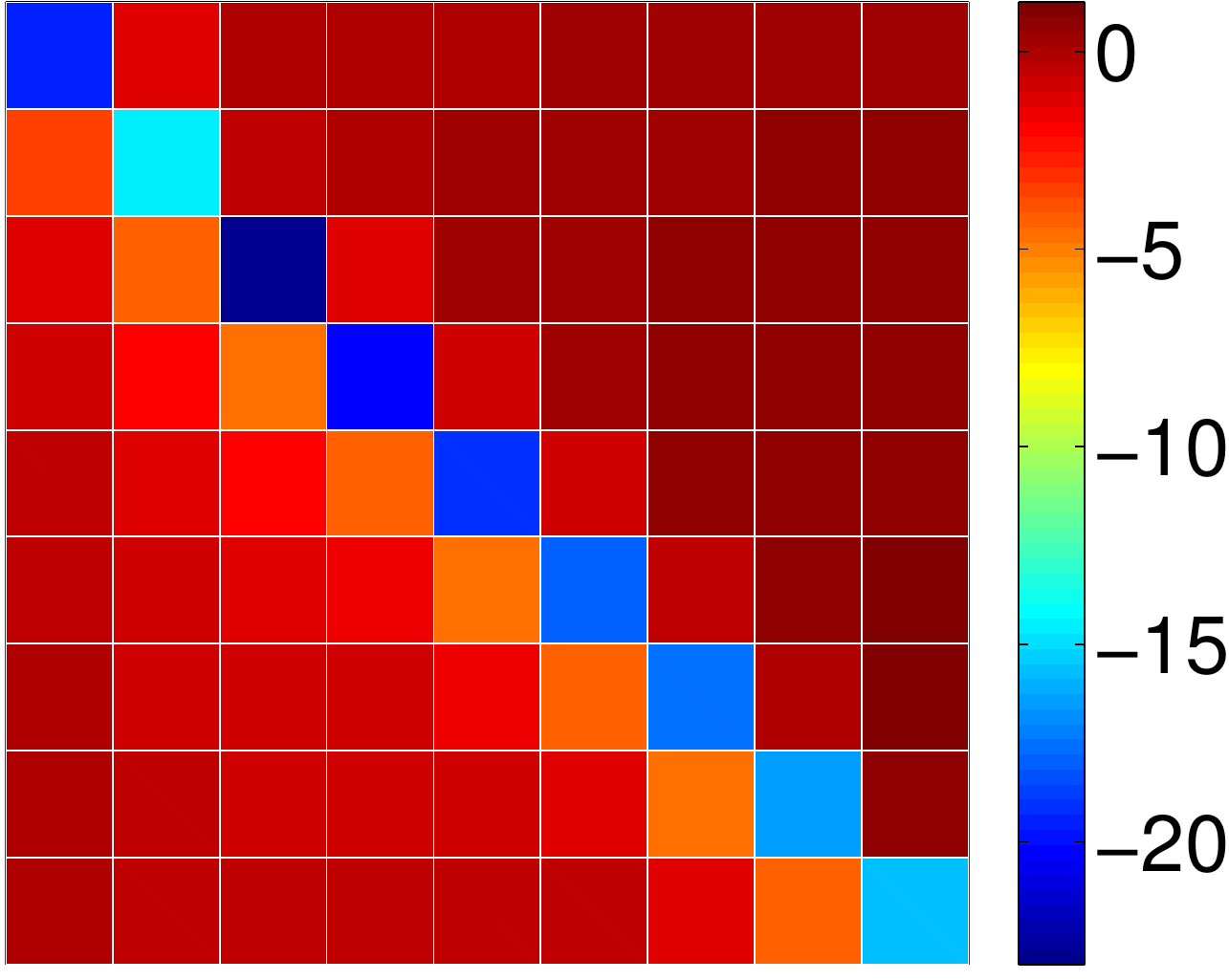}} \qquad
  \subfloat[$\partial  a    /\partial \theta$]{\incfig[width=0.45\columnwidth]{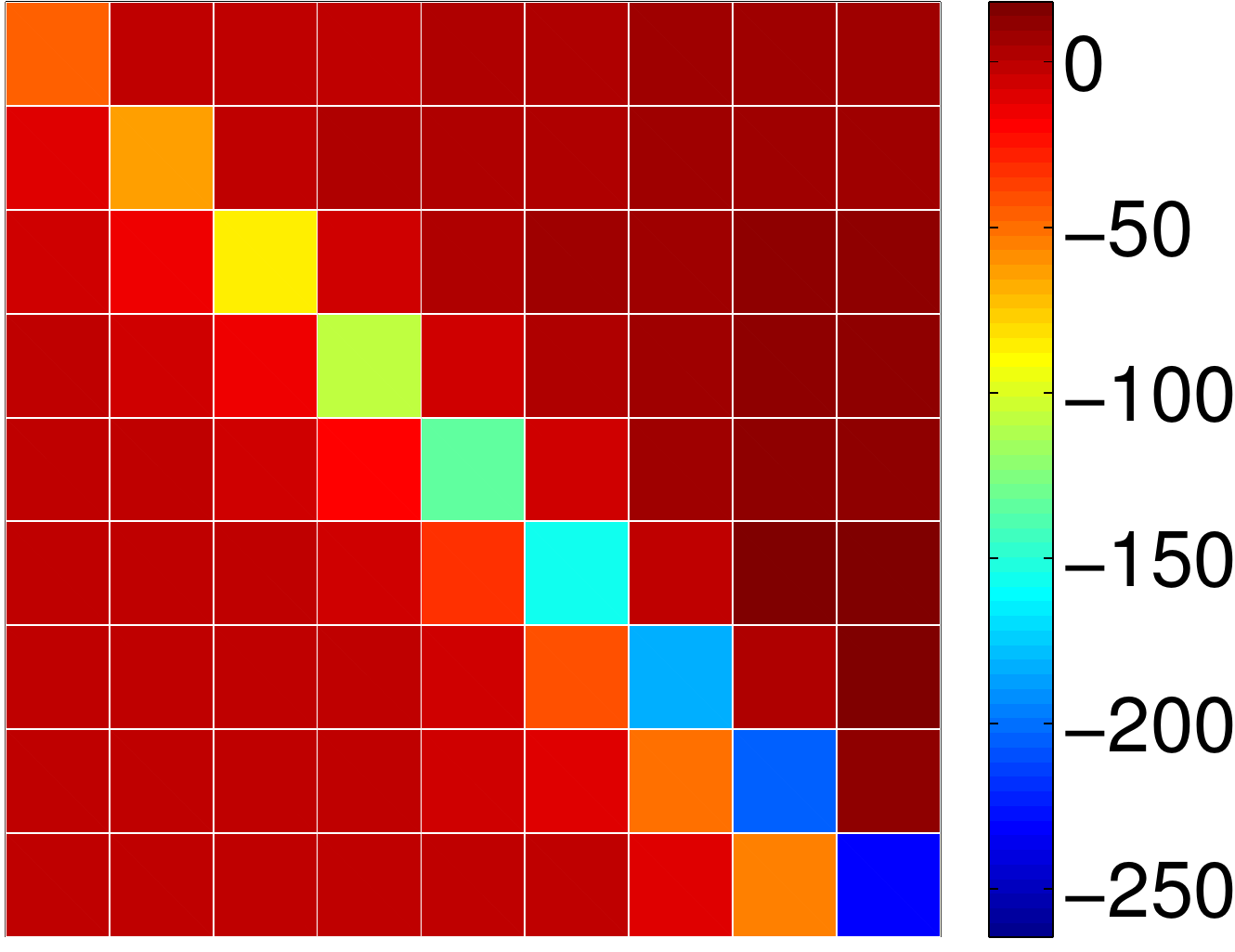}} \qquad
  \caption{Visualization of the Jacobian matrix: (a) a schematic showing the
    arrangement of the four different blocks in the Jacobian matrix, (b,c,d, \&
    e) contour plots for each of the four blocks of the matrix. The contour levels
    are different in each block.}
  \label{fig:jacobian}
\end{figure}

The elements of the Jacobian matrix are calculated using a forward finite
difference formula. Figure~\ref{fig:jacobian} plots the Jacobian matrix, split
in four blocks. The blade is discretized into $m=10$ segments, hence each block
has $10 \times 10$ elements. Each block of the Jacobian matrix in
Fig.~\ref{fig:jacobian} is expected to be diagonally dominant, as a change in
geometry at a given radial location has maximum effect on the aerodynamic
performance at the same location. The effect can be felt at nearby radial
stations as well (off-diagonal terms), but it is much smaller than the diagonal
terms. The results in Fig.~\ref{fig:jacobian} validate this hypothesis. It
should be noted that due to the non-linearity of the function $\bf{F}({\bf x})$
in Eq.~(\ref{e:function}), the Jacobian matrix needs to be updated at each
optimization iteration. Figure.~\ref{fig:jacobian} plots the Jacobian matrix at
the last iteration when convergence is achieved.

\begin{figure}[htb!]
  \subfloat[At first iteration]{\incfig[width=1\columnwidth]{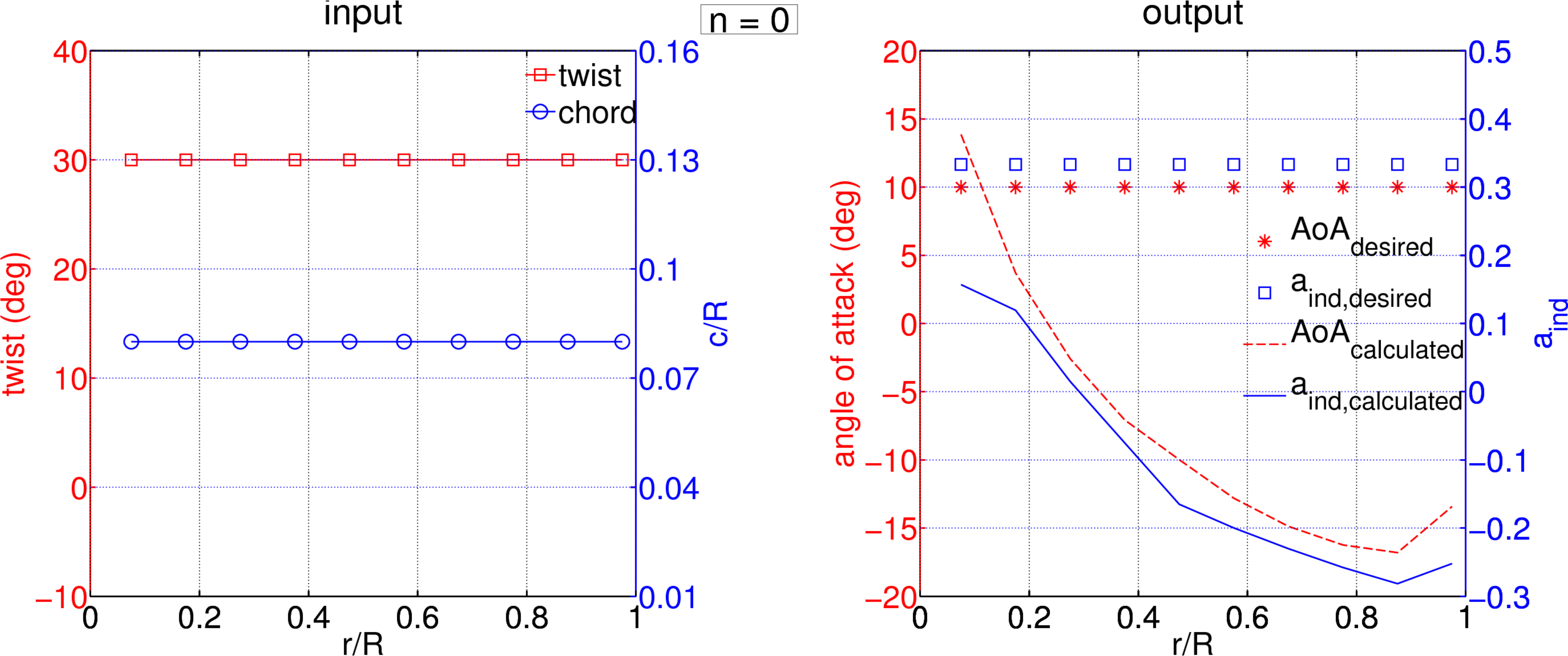}} \\
  \subfloat[At final iteration]{\incfig[width=1\columnwidth]{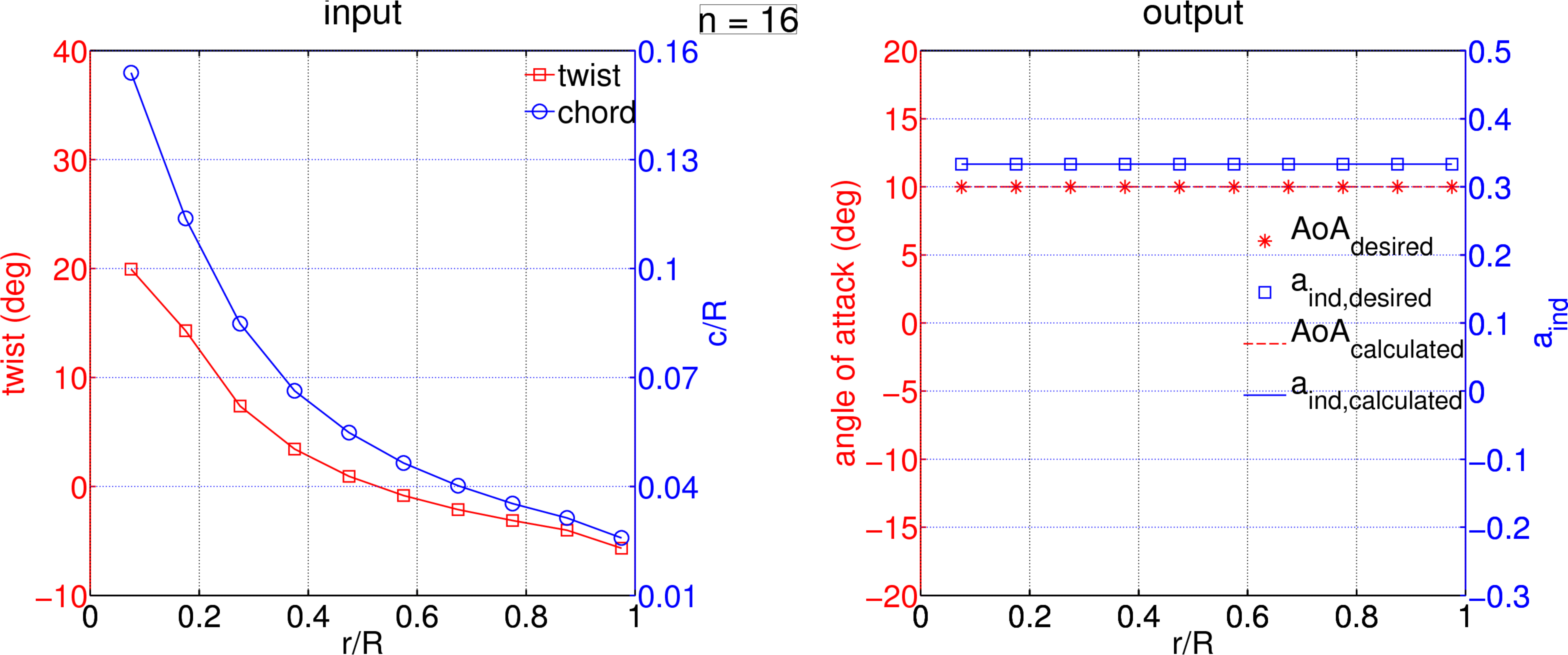}}
  \caption{Results for Test Case $1$: input geometry ($c$ and $\theta$
  distributions) and output aerodynamic performance ($\alpha$ and $a$
  distributions) at the first iteration (top two plots) and final iteration
  (bottom two plots).}%
  \label{fig:test1}%
\end{figure}

Figure~\ref{fig:test1} shows the input and output of the optimization
algorithm. Radius and chord are non-dimensionalized by the tip radius $(R)$.
The initial estimates for both chord and twist are uniform everywhere and are
far from the final distribution. The method converges to the desired values of
$\alpha$ and $a$. The $c$ and $\theta$ distributions of the converged result
(see Fig.~\ref{fig:test1}) are typical of wind turbine blades and the
calculated values of $a$ and $\alpha$ are almost identical to the desired
(prescribed) values (see Fig.~\ref{fig:test1}).

The algorithm was tested with various initial input distributions of $c$ and
$\theta$ and was found to always converge, demonstrating that the method is
robust and insensitive to initial estimates. This test case validates the
capability of the algorithm to perform inverse design of single rotor wind
turbines.

\subsection*{Test Case $2$: SRWT with $\{\alpha,a\}$ Prescribed using RANS/ADM}
\label{sec:test2}
%
The objective of this test case is to verify the inverse algorithm for a
turbine blade that is made of different airfoils along its span. The NREL 5 MW
wind turbine~\cite{jonkman2009definition} is considered for this test. This
turbine rotor blade has cylindrical cross-section at its root and the rest of
the blade is made of seven different airfoils. For this test case, the turbine
geometry as given in Ref.~\cite{jonkman2009definition} is used with the
RANS/ADM solver to obtain $\alpha$ and $a$ distributions. These distributions
are then prescribed as the desired values for the inverse design algorithm.
Constant $c$ and $\theta$ along the blade are used as the initial guess of the
blade geometry. The test of the inverse algorithm is in obtaining the original
$c$ and $\theta$ distributions of Ref.~\cite{jonkman2009definition}.

\begin{figure}[htb!]
  \incfig[width=1\columnwidth]{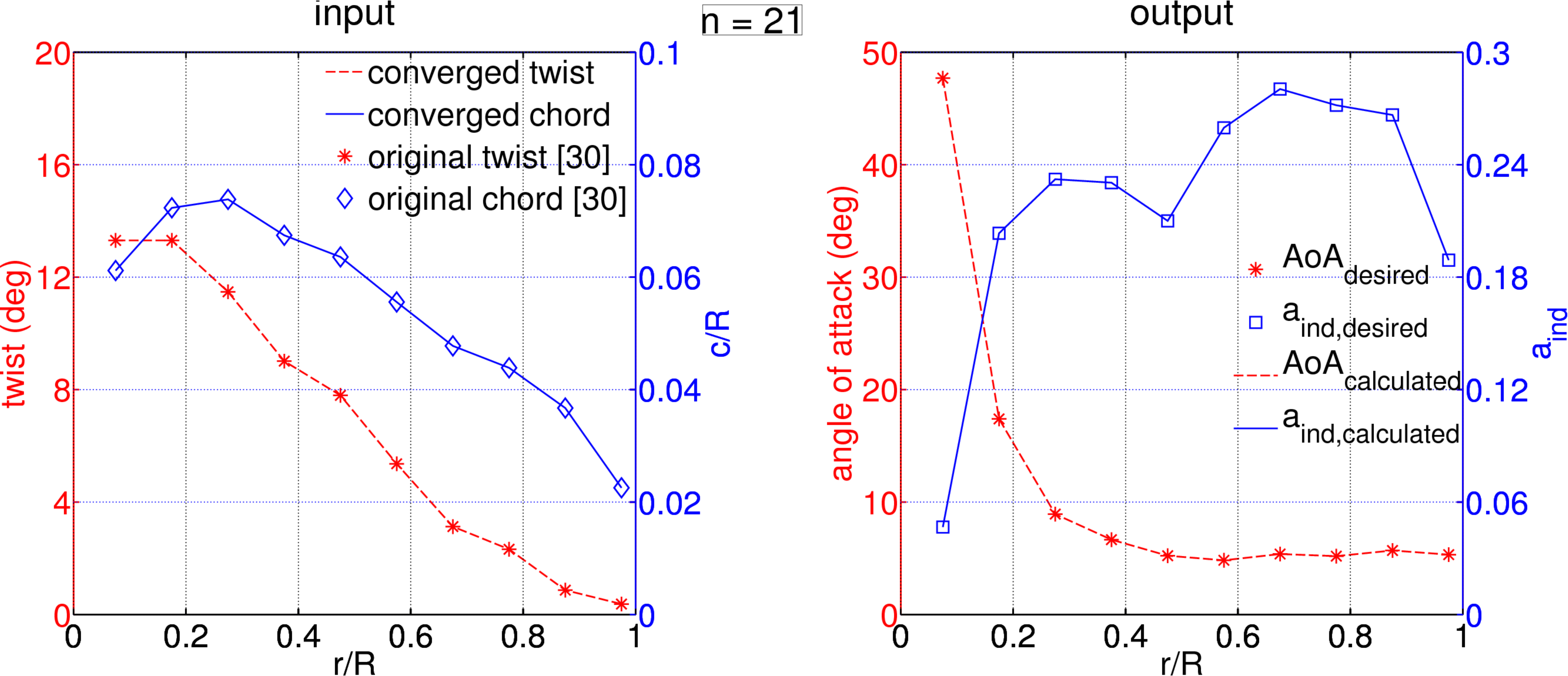}
  \caption{Results for Test Case $2$: Lines show converged geometry ($c$ and
    $\theta$) and converged $\alpha$ and $a$ distributions at the final
    iteration; symbols show original geometry from
    Ref.~\cite{jonkman2009definition} and its aerodynamic performance ($\alpha$
    and $a$) obtained via direct analysis using RANS/ADM.}%
  \label{fig:test2}%
\end{figure}

The inverse algorithm can successfully reproduce the $c$ and $\theta$
distributions with less than $1\%$ error in $\bm{l}^2$ norm after only $21$
iterations. Figure~\ref{fig:test2} shows the original $c$ and $\theta$
distributions as well as the converged results, which are nearly overlaid. The
initial guess of uniform $c$ and $\theta$ distributions are omitted from
Fig.~\ref{fig:test2} for clarity.

\subsection*{Test Case $3$: SRWT with $\{\alpha,a\}$ Prescribed using BEM}
\label{sec:test3}

The proposed inverse design process is applied to a conventional SRWT made
entirely from a single airfoil (DU-96-W180). The test is similar to Test Case
$2$ with the exception that BEM is used as the direct solver instead of
RANS/ADM to obtain the desired distributions of $\alpha$ and $a$. Note that the
inverse design algorithm remains unchanged and it still uses RANS/ADM as its
direct solver.

Due to the differences between RANS/ADM and BEM algorithms, the blade design
obtained using the inverse method cannot be expected to yield exactly the
original geometry ($c$ and $\theta$ distributions) as was the case in Test Case
2. The purpose of this test is to ensure the capability of the presented
inverse design algorithm to yield a geometric design that is consistent with
the direct solver used; in this case RANS/ADM when the desired output is
prescribed by another direct solver.

\begin{figure}[htb!]
  \incfig[width=\columnwidth]{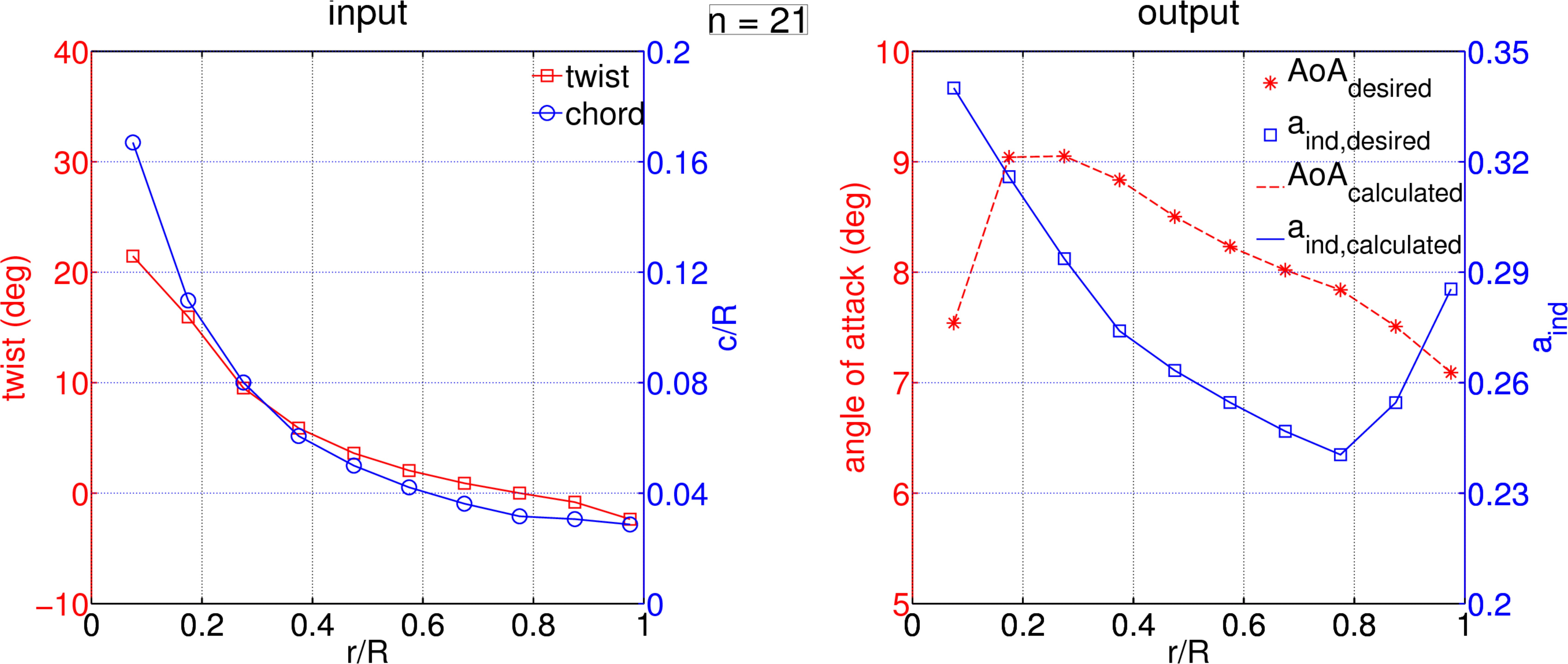}
  \caption{Results for Test Case $3$: Converged $c$ and $\theta$ distributions
    (left) and $\alpha$ and $a$ distributions (right) at the final iteration.
    Predicted $\alpha$ and $a$ distributions (lines) are compared with desired
    values (symbols) in th right plot to show convergence.}%
  \label{fig:test3}%
\end{figure}

Figure~\ref{fig:test3} shows the final results for this test. While the
uniform initial distributions for $c$ and $\theta$ are far from the final
design, the inverse design algorithm can successfully yield a blade geometry
($c$ and $\theta$ distributions) that results in the desired distributions of
$\alpha$ and $a$ over the blade span.

\subsection*{Trust-Region-Reflective Method versus Multi-dimensional Newton Iteration}
\label{sec:OptComparisons}
%
The Trust-Region-Reflective (TRF) method is compared against the
multi-dimensional Newton iteration method for test cases $1$ and $3$. The
multi-dimensional Newton iteration method is a standard optimization method
that has been used in other wind turbine inverse design approaches (see e.g.,
Refs.~\cite{selig1995development,lee2015inverse}). The variation with iteration
number of the $\bm{l}^2$ norm of the objective function $\bm{F}$, which is
representative of residual error, is compared between the two methods. It
should be noted that in order to ensure a physically realistic blade shape, the
blade chord is constrained to have a value between $0.01\,R$ to $0.2\,R$.

\begin{figure}[htb!]
  \incfig[width=0.55\columnwidth]{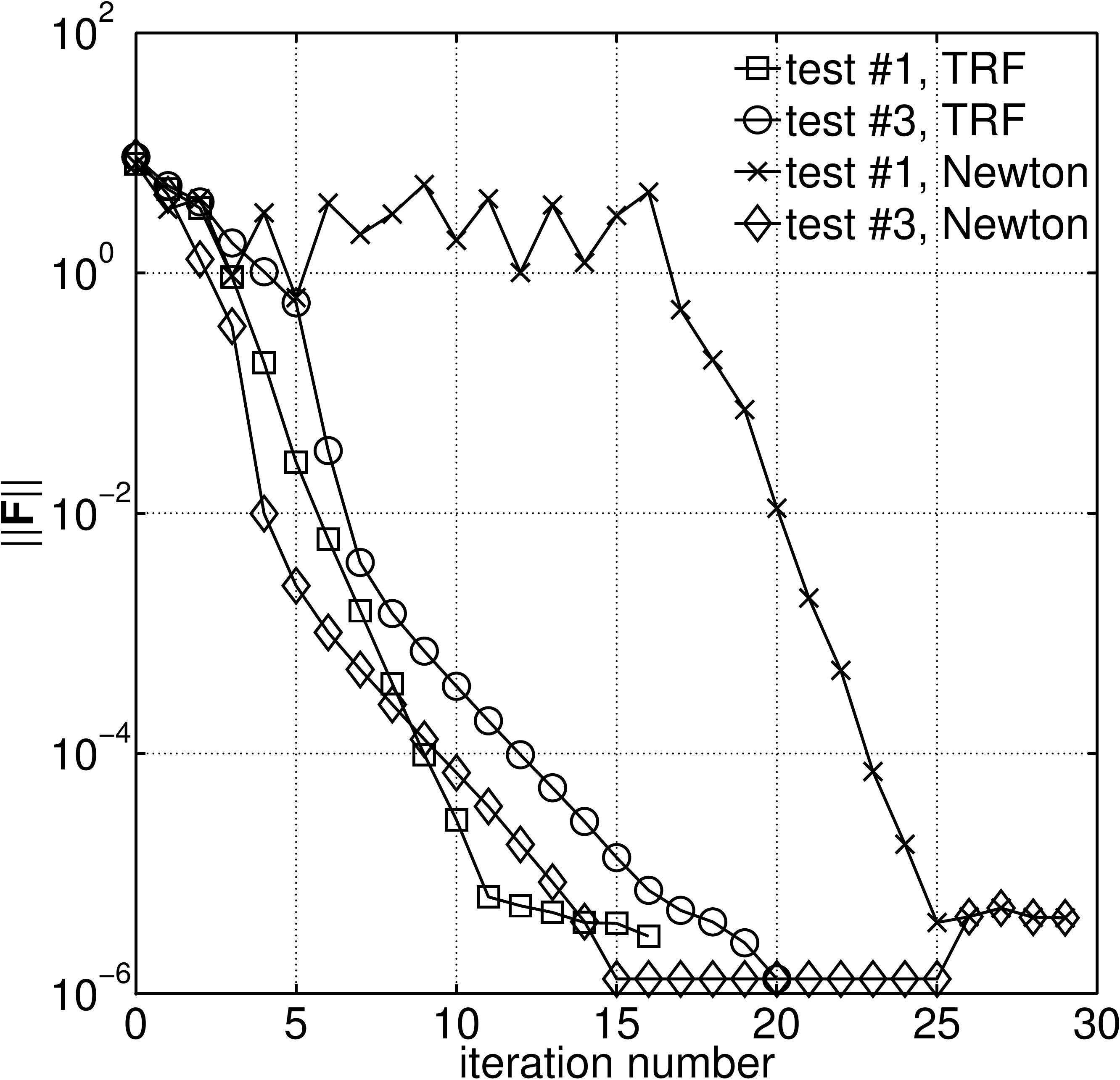}
  \caption{Comparison between multi-dimensional Newton iteration and
    Trust-Region-Reflective (TRF) optimization methods.}
  \label{fig:NewtonVsTRF}%
\end{figure}

As demonstrated in Fig.~\ref{fig:NewtonVsTRF}, the decrease in the norm of
$\bm{F}$ for Test Case $1$ is much faster with the TRF
method than with the multi-dimensional Newton iteration method. In this case,
the Newton method is trying to push the blade chord at some radial stations
beyond the specified constraint of $0.2\,R$. To impose the constraint, the
chord is limited to $0.2\,R$ at each iteration. This results in oscillations in
the $\bm{l}^2$ norm of $\bm{F}$.

When the chord values do not hit a constraint during convergence, as is the
case in Test Case $3$, the Newton method converges faster than the
TRF method. In general, the Newton method converges faster
for {\em unconstrained} problems while the TRF method is
found to be more stable in {\em constrained} blade design problems. The reason
for this additional stability of the TRF method can be
attributed to the gradual expansion of the trusted region in which the solution
to the sub-problem is sought (see Section~\ref{sec:inverse}). Nonetheless, both
methods converge, and ultimately yield identical $c$ and $\theta$
distributions.

\section*{Extension to Multi-Rotor Wind Turbines}
\label{sec:MRWT}
%
This section presents the extension of the inverse blade design process to
dual-rotor wind turbines (DRWTs). The procedure can be easily extended to
multi-rotor turbines with more than two rotors by applying the same idea
presented here.

\begin{figure}[htb!]
  \centering
  {\incfig[width=0.4\columnwidth]{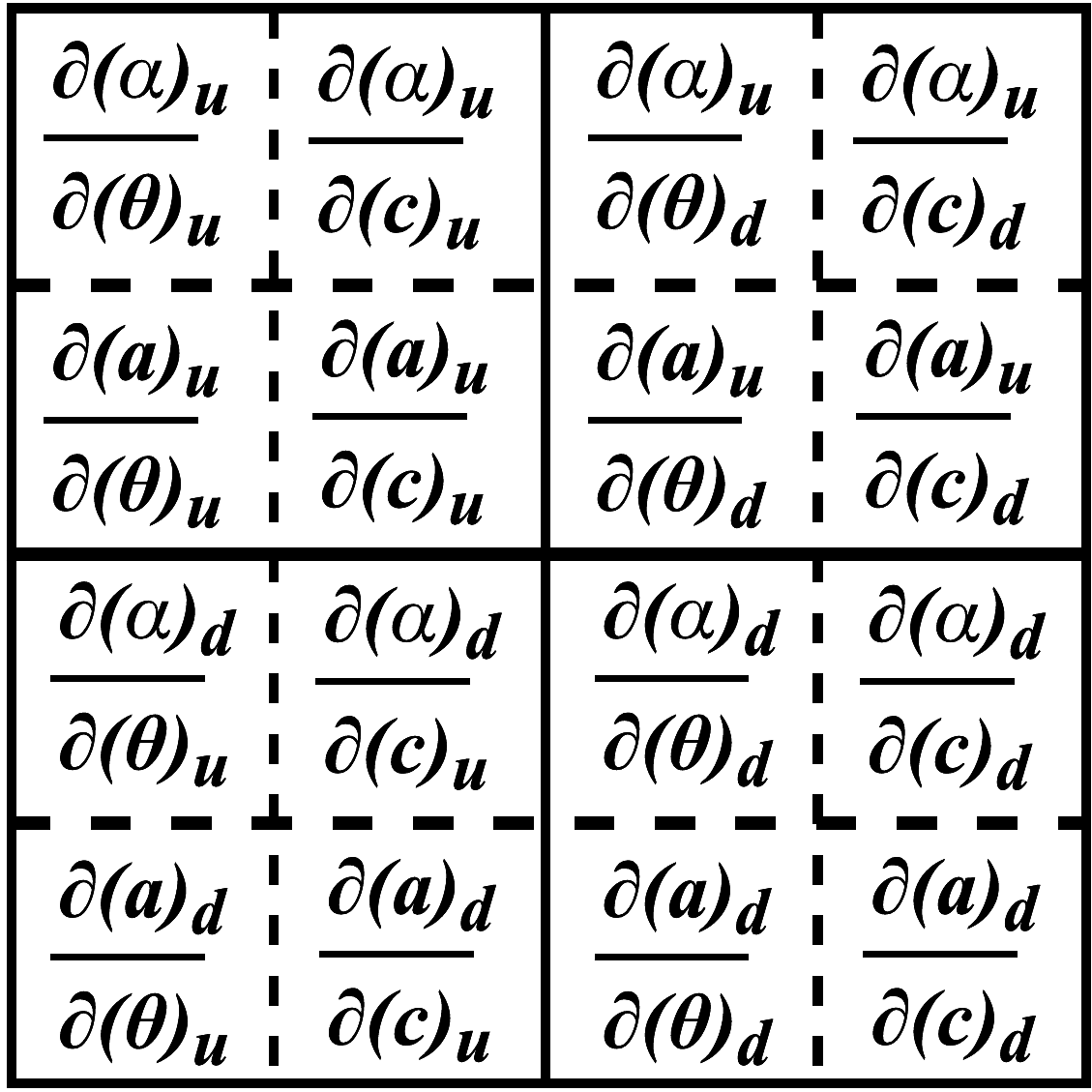}}
  \caption{Schematic of the Jacobian for DRWT cases.}
  \label{fig:JacobianDRWT}%
\end{figure}

For inverse design of DRWTs, modifications in computational set-up as well as
arrangement of the Jacobian matrix are made. As mentioned in
Section~\ref{sec:intro}, DRWTs use a smaller, secondary rotor upstream of the
main rotor in order to reduce blade root loss and enhance momentum entrainment.
Two cases of DRWTs with different blades and rotor radius ratio, $R_u/R_d$ are
presented in this section where subscripts $u$ and $d$ stand for upstream and
downstream rotors, respectively.  Upstream rotor sits at $x_u=-Sep/2$ and
downstream rotor is located at $x_d=+Sep/2$ where $Sep$ is the rotor-rotor
separation distance between the two rotors. Both rotors are modeled as actuator
disks and the mesh is refined in the vicinity of the two rotor disk locations
in both $x$ and $z$ directions.  Independent variables are $\theta$ and $c$
distributions, and target parameters are radial profiles of $\alpha$ and $a$
for both rotors. For DRWTs, the Jacobian (sensitivity) matrix not only includes
the effects of changing design variables of each rotor on its own target
variables, but also interaction effects between the rotors. A schematic of the
Jacobian matrix is shown in Fig.~\ref{fig:JacobianDRWT}. For DRWTs, it has 4
quadrants and each quadrant has 4 blocks. The diagonal quadrants model
self-influence of each rotor, while the off-diagonal quadrants model
rotor-rotor interaction effects.

Aerodynamic interactions (coupling) between the two rotors of a DRWT makes the
design process more complicated. The main interaction between the two rotors is
due to the fact that a part of the downstream rotor operates in the wake of the
upstream rotor. A change in the geometry of the upstream rotor affects its
wake, and  for practical rotor-rotor separation distances (small compared to
rotor radius), the aerodynamic response of the downstream rotor. The downstream
rotor also has a potential field which affects the aerodynamics of the upstream
rotor, although this effect is expected to be much smaller compared to that due to
the wake. The Jacobian matrix is helpful in visualizing, understanding, and
quantifying such interaction effects.

To verify the extension of the inverse design algorithm to DRWTs, several cases
corresponding to the SRWT test cases $2$ and $3$ were attempted.  The inverse
algorithm was able to successfully obtain the original distributions of $c$ and
$\theta$ in all cases. Only two of these cases are reported here for brevity.

\subsection*{Test Case $4$: Inverse Design of DRWTs}
\label{sec:test4}

\begin{figure}[htb!]
  \incfig[width=1\columnwidth]{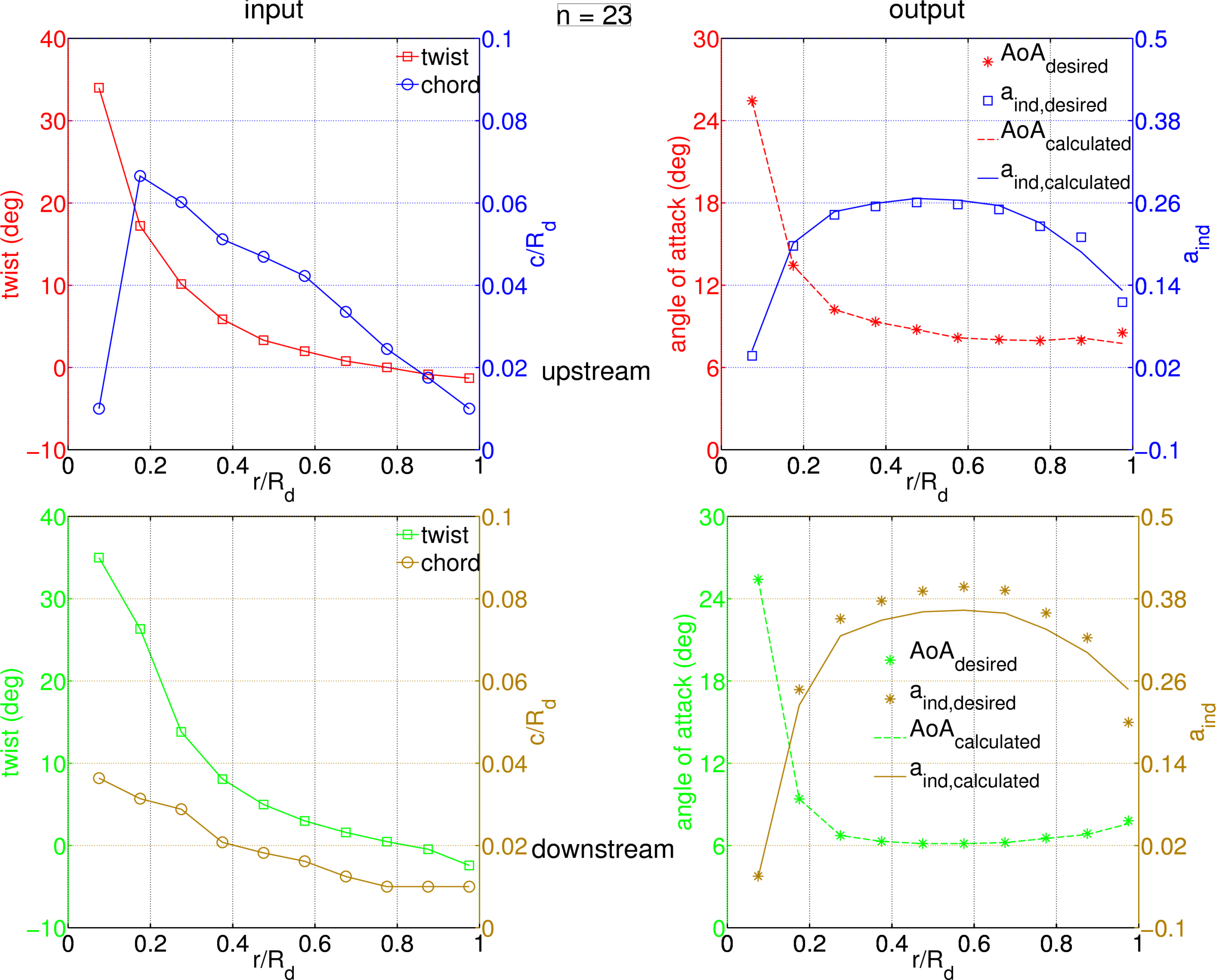}
  \caption{Results for Test Case $4$: Converged geometry ($c$ and $\theta$) and
    aerodynamics ($\alpha$ and $a$) for upstream (top two plots) and downstream
    (bottom two plots) rotors. Plots on the right show desired aerodynamic
    performance with symbols and predicted performance of the final geometry
    with lines.}
  \label{fig:test4}%
\end{figure}

The DRWT considered here has two equal-size rotors ($R_u/R_d=1$), both made
entirely of the DU-96-W180 airfoil with rotor-rotor separation of $0.3\,R_d$
and tip speed ratio $\lambda=7.0$. Similar to test case $3$, a different
aerodynamic analysis solver is used to obtain radial profiles of $\alpha$ and
$a$, which are then prescribed as the desired performance outcomes. The vortex
lattice method (VLM) proposed by Rosenberg and
Sharma~\cite{rosenberg2016prescribed} is selected as the direct solver to
analyze the DRWT. The objective is to see if the inverse solver is able to
obtain a design that gives the desired aerodynamic performance (in this case,
obtained using another wind turbine analysis software). The final results of
the inverse design are shown in Fig.~\ref{fig:test4}. The inverse design
algorithm is able to obtain geometries for both rotors of the DRWT that nearly
satisfy the prescribed aerodynamic performance; there is a little difference
between the desired and calculated aerodynamic performance, particularly axial
induction, at the final iteration of the inverse algorithm. This difference is
due to the constraint imposed on the blade chord; chord values are restricted
at all radial locations for each rotor to be between $0.01$ to $0.2$ times the
tip radius of the corresponding rotor.  In this test case, chord values near
the tip of the downstream rotor reach the lower limit ($0.01\,R_d$ in
Fig.~\ref{fig:test4}, bottom left) and this restricts the optimization
algorithm from identically reproducing the target/desired values.

%
\subsection{Test Case $5$: Inverse Design of DRWTs}
\label{sec:test5}

The aim of this test case is to design a more realistic DRWT which has two
rotors of different sizes, and the downstream rotor has different airfoils
along its span. The upstream rotor is a Betz-optimum rotor, which uses only the
DU-96-W180 airfoil and the downstream rotor uses the NREL Phase VI turbine
rotor blades~\cite{hand2001unsteady}. The NREL Phase VI turbine rotor uses the
S809 airfoil from $r/R \geq 0.25$ to the tip, and cylinder at its blade root
region; transition sections are not included. The 2D polar data of the S809
airfoil at chord-based Reynolds number, $Re_c=5\times 10^5$ were obtained using
XFOIL~\cite{drela1989xfoil}. Separation between the rotors is $0.3\,R_d$ with
$R_u/R_d=0.3$ and  the tip speed ratio of either rotor, defined using its tip rotor radius,
is $7.0$. A test case similar to Test Case $2$ is performed, in which
the direct solver (RANS/ADM) is evoked with a known geometry ($c$ and $\theta$
distributions for both rotors). The RANS/ADM computed $\alpha$ and $a$
distributions for both rotor blades are then set as target values in the design
process.

\begin{figure}[htb!]
  \incfig[width=\columnwidth]{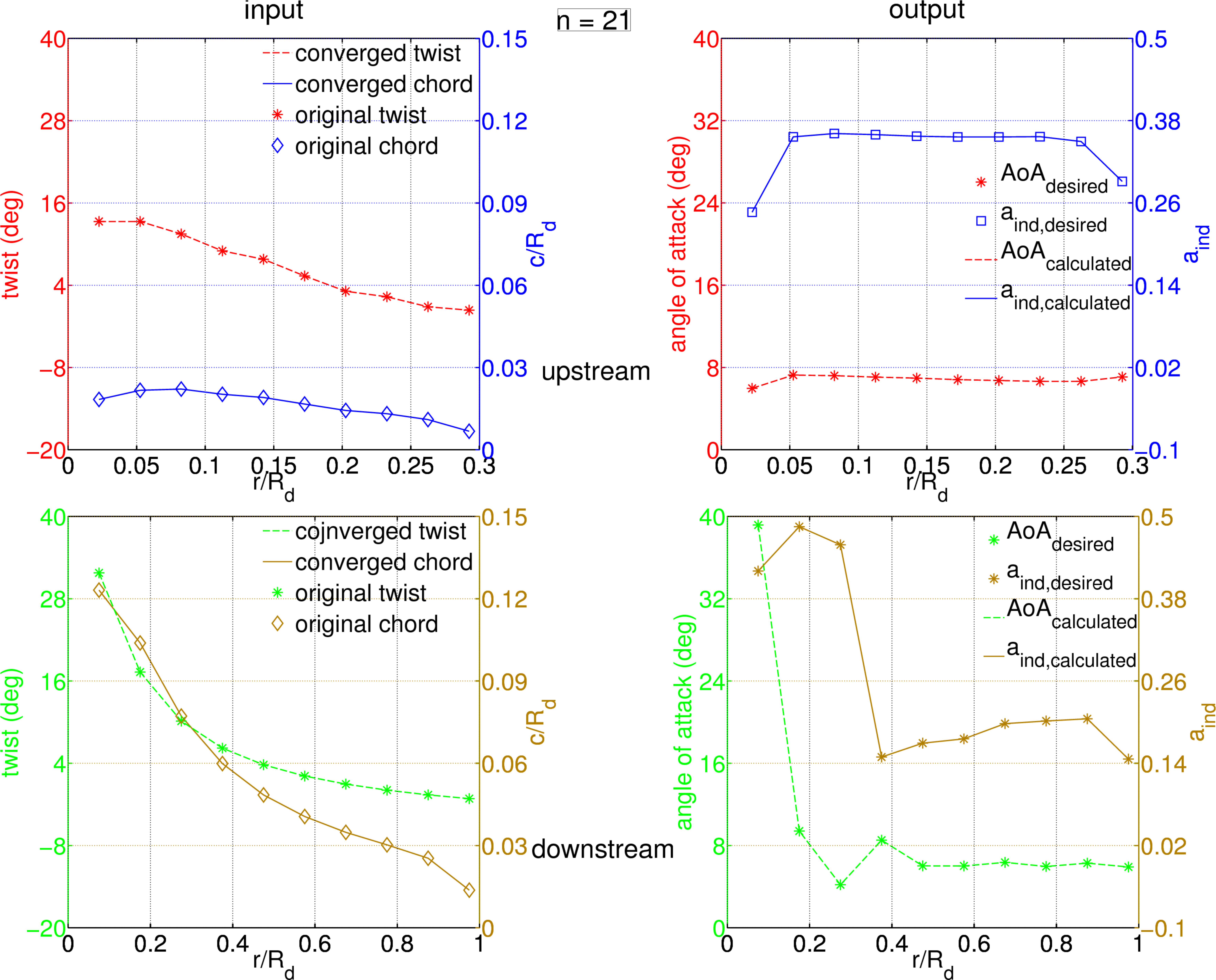}
  \caption{Results for Test Case $5$: Converged geometry ($c$ and $\theta$) and
  corresponding $\alpha$ and $a$ distributions for upstream (top two plots) and
  downstream (bottom two plots) rotors. Symbols show original $c,\theta$ and
  $\alpha,a$, and lines show corresponding converged quantities.}%
  \label{fig:test5}%
\end{figure}

Initial estimates of $c$ and $\theta$ for both rotors are constant values
($c_u=0.03\,R_d$, $c_d=0.07\,R_d$ and $\theta_u=\theta_d=35^\circ$) along the
span.  The converged geometry ($c$ and $\theta$) and aerodynamic performance
($\alpha$ and $a$) are presented in Fig.~\ref{fig:test5}. The desired
distributions of $\alpha$ and $a$ are achieved through inverse design, and the
converged geometry is identical to the original geometry.

\begin{figure*}[htb!]
    \centering
    \incfig[width=1.0\columnwidth]{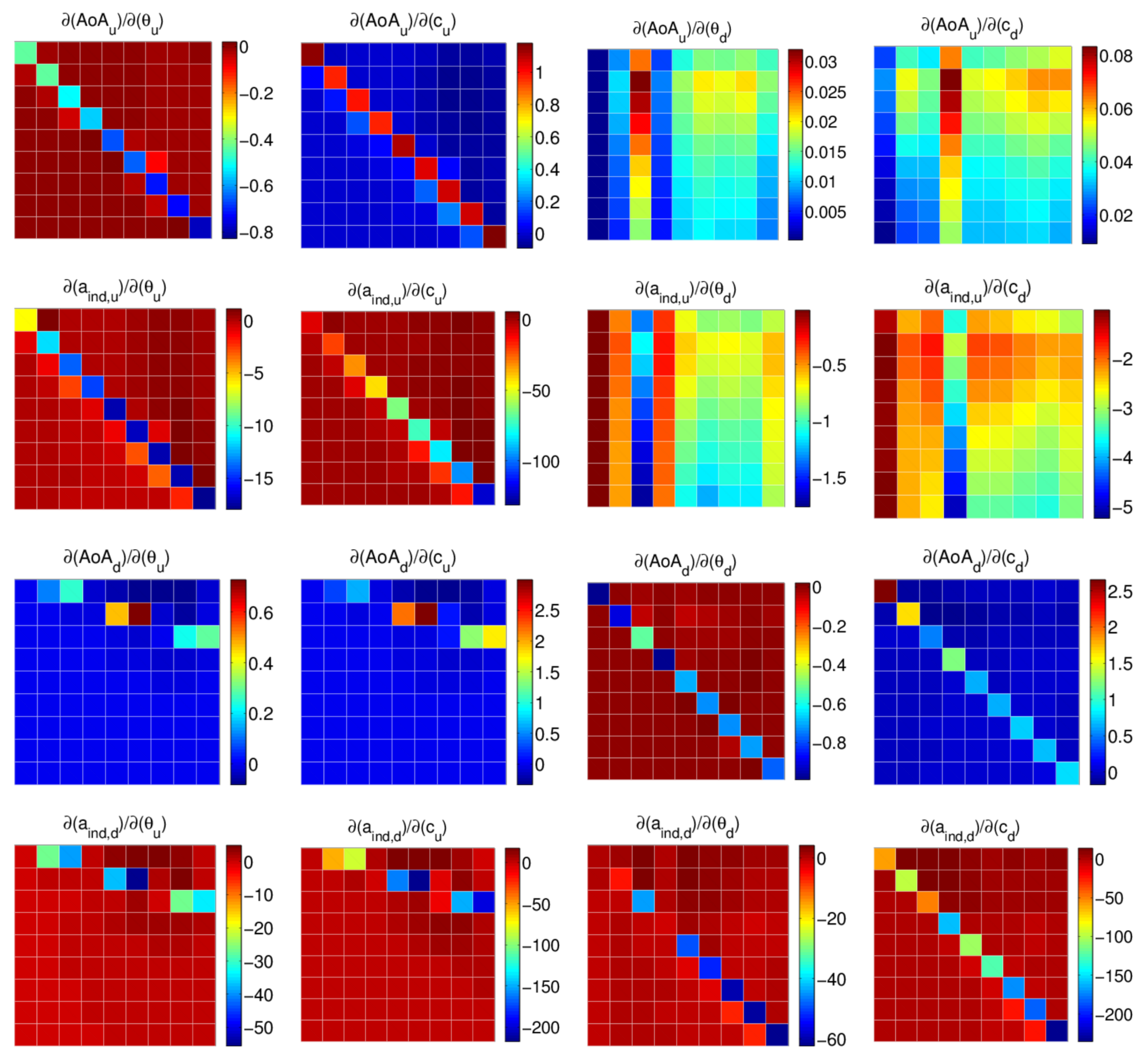} \,
    \caption{Different blocks of the Jacobian matrix for Test Case $5$ after
    convergence.}%
    \label{fig:test5Jacobi}%
\end{figure*}

The Jacobian at the last iteration is shown in Fig.~\ref{fig:test5Jacobi}. As
expected, the blocks in the diagonal (upper left and bottom right) quadrants
are strongly diagonal because they represent the sensitivity of output parameters to
the inputs of the same rotor (refer to the schematic in
Fig.~\ref{fig:JacobianDRWT}). The upper right quadrant shows the effect of
input parameters of the downstream rotor on the output of the upstream rotor.
This effect is almost negligible because the rotor-rotor separation distance
($=0.3\,R_d$) is too large for the potential field of the downstream rotor to
substantially influence the upstream rotor aerodynamics. It should be noted
that the only potential field captured here is due to the aerodynamic pressure;
thickness effects are not modeled. Rosenberg and
Sharma~\cite{rosenberg2016prescribed} has shown that thickness effects can be
neglected at such separation distances.

Lastly, the bottom left quadrant has non-zero components up to the tip radius
of the upstream rotor ($R_u/R_d=0.3$) since any change in the shape of the
smaller upstream rotor will change the flow speed and angle in its wake and
eventually the downstream rotor will be affected over a partial span. The
downstream blade sections with $r>R_u$ $(= 0.3\,R_d$ here) will not experience
much difference by changing the geometry of the upstream rotor.

\section*{Conclusion}
\label{sec:conclusion}

An inverse design algorithm for wind turbine blade design is presented in this
paper. The goal is to realize the desired aerodynamic performance by
iteratively changing the blade geometry until convergence. The design
parameters are axial induction factor and angle of attack, while the
independent variables are dimensionless chord and twist distributions along the
blade. The inverse design procedure requires the flow field to be solved around
the wind turbine blades. In this study, RANS/ADM is chosen as the direct solver
to compute wind turbine aerodynamic performance while the
Trust-Region-reflective (TRF) algorithm is selected as the iterative searching
method.

The design algorithm is tested with different single- and dual-rotor wind
turbines. The purpose of each case is discussed and then the algorithm is put
to the test to demonstrate its ability for accomplishing the design goal. In
order to achieve a realistic design, constraints are applied to chord values.
It is seen that when this constraint is not active (i.e. all cases except Test
Case $4$), the proposed inverse design algorithm converges to the blade
geometry that yields the desired aerodynamic performance. However, there is a
small difference between the desired and calculated values in Test Case $4$
because the applied constraint does not let the procedure drive the design to
minimize residual error.

The TRF method is compared to the multi-dimensional Newton iteration method. It
is found that while both methods are capable of handling the inverse problem,
and converge to the same geometric distributions, they have different
convergence rates for different test cases. When the design constraints come in
play, the Newton method exhibits oscillations and its convergence is poor.
However, when the constraints are not active, as in Test Case $3$, the Newton
method converges faster than TRF. In general, the TRF method was found to be
more stable for constrained problems.

Another objective of this study is to extend the blade design process to
multi-rotor wind turbines (DRWTs). The analysis presented in this paper
demonstrates the ability of the inverse design algorithm to obtain blade
geometries that satisfy the desired aerodynamic performance of DRWTs. The
differences in SRWT and DRWT Jacobian matrices were discussed. The Jacobian
matrix quantifies the sensitivity of the output to the input parameters.  For
DRWTs, it also demonstrates the aerodynamic coupling between the two rotors.
The geometry of the downstream rotor blade is found to have a negligible effect
on the upstream rotor for relatively large rotor-rotor separation distances.
The upstream rotor however has a considerable effect on the aerodynamic
performance of the downstream rotor at radial locations affected by the wake of
the upstream rotor.

\section*{Acknowledgments}
\label{sec:ack}
Funding for this work was provided by the National Science Foundation under
grant number NSF/CBET-1438099. The numerical simulations reported here were
performed using the NSF XSEDE resources available to the authors under grant
number TG-CTS130004, and the Argonne Leadership Computing Facility, which is a
DOE Office of Science User Facility supported under Contract DE-AC02-06CH11357.


\bibliography{references}

\begin{thebibliography}{10}

\bibitem{adkins1994design}
C.~N. Adkins and R.~H. Liebeck.
\newblock Design of optimum propellers.
\newblock {\em Journal of Propulsion and Power}, 10(5):676--682, 1994.

\bibitem{alaimo20153d}
A.~Alaimo, A.~Esposito, A.~Messineo, C.~Orlando, and D.~Tumino.
\newblock 3d cfd analysis of a vertical axis wind turbine.
\newblock {\em Energies}, 8(4):3013--3033, 2015.

\bibitem{bachant2016modeling}
P.~Bachant and M.~Wosnik.
\newblock Modeling the near-wake of a vertical-axis cross-flow turbine with 2-d
  and 3-d rans.
\newblock {\em arXiv preprint arXiv:1604.02611}, 2016.

\bibitem{byrd2000trust}
R.~H. Byrd, J.~C. Gilbert, and J.~Nocedal.
\newblock A trust region method based on interior point techniques for
  nonlinear programming.
\newblock {\em Mathematical Programming}, 89(1):149--185, 2000.

\bibitem{byrd1987trust}
R.~H. Byrd, R.~B. Schnabel, and G.~A. Shultz.
\newblock A trust region algorithm for nonlinearly constrained optimization.
\newblock {\em SIAM Journal on Numerical Analysis}, 24(5):1152--1170, 1987.

\bibitem{chattot2003optimization}
J.-J. Chattot.
\newblock Optimization of wind turbines using helicoidal vortex model.
\newblock {\em Journal of solar energy engineering}, 125(4):418--424, 2003.

\bibitem{coleman1994convergence}
T.~F. Coleman and Y.~Li.
\newblock On the convergence of interior-reflective newton methods for
  nonlinear minimization subject to bounds.
\newblock {\em Mathematical programming}, 67(1-3):189--224, 1994.

\bibitem{drela1989xfoil}
M.~Drela.
\newblock Xfoil: An analysis and design system for low reynolds number
  airfoils.
\newblock In {\em Low Reynolds number aerodynamics}, pages 1--12. Springer,
  1989.

\bibitem{giguere1997desirable}
P.~Giguere and M.~Selig.
\newblock Desirable airfoil characteristics for large variable-speed horizontal
  axis wind turbines.
\newblock {\em Journal of solar energy engineering}, 119(3):253--260, 1997.

\bibitem{goldstein1929vortex}
S.~Goldstein.
\newblock On the vortex theory of screw propellers.
\newblock {\em Proceedings of the Royal Society of London. Series A, Containing
  Papers of a Mathematical and Physical Character}, 123(792):440--465, 1929.

\bibitem{hand2001unsteady}
M.~Hand, D.~Simms, L.~Fingersh, D.~Jager, J.~Cotrell, S.~Schreck, and
  S.~Larwood.
\newblock Unsteady aerodynamics experiment phase vi: wind tunnel test
  configurations and available data campaigns.
\newblock {\em National Renewable Energy Laboratory, Golden, CO, Report No.
  NREL/TP-500-29955}, 2001.

\bibitem{hargreaves2007use}
D.~Hargreaves and N.~G. Wright.
\newblock On the use of the k--$\varepsilon$ model in commercial cfd software
  to model the neutral atmospheric boundary layer.
\newblock {\em Journal of Wind Engineering and Industrial Aerodynamics},
  95(5):355--369, 2007.

\bibitem{jonkman2009definition}
J.~Jonkman, S.~Butterfield, W.~Musial, and G.~Scott.
\newblock Definition of a 5-mw reference wind turbine for offshore system
  development.
\newblock {\em National Renewable Energy Laboratory, Golden, CO, Technical
  Report No. NREL/TP-500-38060}, 2009.

\bibitem{lam2016study}
H.~Lam and H.~Peng.
\newblock Study of wake characteristics of a vertical axis wind turbine by
  two-and three-dimensional computational fluid dynamics simulations.
\newblock {\em Renewable Energy}, 90:386--398, 2016.

\bibitem{launder1974numerical}
B.~E. Launder and D.~Spalding.
\newblock The numerical computation of turbulent flows.
\newblock {\em Computer methods in applied mechanics and engineering},
  3(2):269--289, 1974.

\bibitem{lee2010two}
K.-H. Lee, K.-H. Kim, D.-H. Lee, K.-T. Lee, and J.-P. Park.
\newblock Two-step optimization for wind turbine blade with probability
  approach.
\newblock {\em Journal of Solar Energy Engineering}, 132(3):034503, 2010.

\bibitem{lee2015inverse}
S.~Lee.
\newblock Inverse design of horizontal axis wind turbine blades using a vortex
  line method.
\newblock {\em Wind Energy}, 18(2):253--266, 2015.

\bibitem{marsh2015three}
P.~Marsh, D.~Ranmuthugala, I.~Penesis, and G.~Thomas.
\newblock Three-dimensional numerical simulations of straight-bladed vertical
  axis tidal turbines investigating power output, torque ripple and mounting
  forces.
\newblock {\em Renewable Energy}, 83:67--77, 2015.

\bibitem{mikkelsen2003actuator}
R.~Mikkelsen.
\newblock {\em Actuator disc methods applied to wind turbines}.
\newblock PhD thesis, Technical University of Denmark, 2003.

\bibitem{moghadassian2015numerical}
B.~Moghadassian, A.~Rosenberg, H.~Hu, and A.~Sharma.
\newblock Numerical investigation of aerodynamic performance and loads of a
  novel dual rotor wind turbine.

\bibitem{mogh2016energies}
B.~Moghadassian, A.~Rosenberg, and A.~Sharma.
\newblock Numerical investigation of aerodynamic performance and loads of a
  novel dual rotor wind turbine.
\newblock {\em Energies}, 9(571), 2016.

\bibitem{patankar1972calculation}
S.~V. Patankar and D.~B. Spalding.
\newblock A calculation procedure for heat, mass and momentum transfer in
  three-dimensional parabolic flows.
\newblock {\em International journal of heat and mass transfer},
  15(10):1787--1806, 1972.

\bibitem{rosenberg2016thesis}
A.~Rosenberg.
\newblock {\em A Computational Analysis of Wind Turbine and Wind Farm
  Aerodynamics with a Focus on Dual Rotor Wind Turbines}.
\newblock PhD thesis, Iowa State University, 2016.

\bibitem{rosenberg2014novel}
A.~Rosenberg, S.~Selvaraj, and A.~Sharma.
\newblock A novel dual-rotor turbine for increased wind energy capture.
\newblock In {\em Journal of Physics: Conference Series}, volume 524, page
  012078. IOP Publishing, 2014.

\bibitem{rosenberg2016prescribed}
A.~Rosenberg and A.~Sharma.
\newblock A prescribed-wake vortex lattice method for preliminary design of
  co-axial, dual-rotor wind turbines.
\newblock {\em Journal of Solar Energy Engineering}, 138(6), 2016.

\bibitem{selig1996application}
M.~S. Selig and V.~L. Coverstone-Carroll.
\newblock Application of a genetic algorithm to wind turbine design.
\newblock {\em Journal of Energy Resources Technology}, 118(1):22--28, 1996.

\bibitem{selig1995development}
M.~S. Selig and J.~L. Tangler.
\newblock Development and application of a multipoint inverse design method for
  horizontal axis wind turbines.
\newblock {\em Wind Engineering}, 19(2):91--106, 1995.

\bibitem{selvaraj2014numerical}
S.~Selvaraj.
\newblock {\em Numerical investigation of wind turbine and wind farm
  aerodynamics}.
\newblock PhD thesis, Iowa State University, 2014.

\bibitem{sorensen1992unsteady}
J.~N. S{\o}rensen and A.~Myken.
\newblock Unsteady actuator disc model for horizontal axis wind turbines.
\newblock {\em Journal of Wind Engineering and Industrial Aerodynamics},
  39(1):139--149, 1992.

\bibitem{thelen2016direct}
A.~S. Thelen, L.~T. Leifsson, A.~Sharma, and S.~Koziel.
\newblock Direct and surrogate-based optimization of dual-rotor wind turbines.
\newblock In {\em 34th Wind Energy Symposium}, page 1265, 2016.

\bibitem{troldberg2008actuator}
N.~Troldberg.
\newblock Actuator line modeling of wind turbine wakes (ph. d. thesis).
\newblock {\em Department of Mechanical Engineering, Technical University of
  Denmark, Lynby}, 2008.

\bibitem{wang2016}
Z.~Wang, W.~Tian, A.~Ozbay, A.~Sharma, and H.~Hui.
\newblock An experimental study on the aeromechanics and wake characteristics
  of a novel twin-rotor wind turbine in a turbulent boundary layer flow.
\newblock {\em Experiments in Fluids}, 57(-):150, 2016.

\end{thebibliography}

\bibliographystyle{abbrv}

\end{document}